\documentstyle [12pt,twoside, epsf]{article}

\def\picill#1by#2(#3)
{\vbox to #2
{\hrule width #1 height 0pt depth 0pt
\vfill\epsffile{#3}}}

\begin{document}

\date{}

\title{\bf Braiding Operators are Universal Quantum Gates}

\author{Louis
H. Kauffman\\ Department of Mathematics, Statistics \\ and Computer Science (m/c
249)    \\ 851 South Morgan Street   \\ University of Illinois at Chicago\\
Chicago, Illinois 60607-7045\\ $<$kauffman@uic.edu$>$\\ and \\ Samuel J. Lomonaco
Jr. \\ Department of Computer Science and Electrical Engineering \\ University of
Maryland Baltimore County \\ 1000 Hilltop Circle, Baltimore, MD 21250\\
$<$lomonaco@umbc.edu$>$}

 \maketitle
  
 \thispagestyle{empty}
 
 \subsection*{\centering Abstract}

{\em
 This paper explores of the role of unitary braiding operators in quantum computing. We show that a single
specific solution $R$ (the Bell basis change matrix) of the Yang-Baxter Equation is a universal gate for quantum computing, in the
presence of local unitary transformations. We show that this same $R$ generates a new non-trivial invariant of braids, knots, and links. 
Other solutions of the Yang-Baxter Equation are also shown to be universal for quantum computation. The paper discusses these
results in the context of comparing quantum and topological points of view. In particular, we discuss quantum computation of link
invariants, the relationship between quantum entanglement and topological entanglement, and the structure of braiding in a topological
quantum field theory.}
 
\section{Introduction}
It is a challenge to unravel the relationships among quantum entanglement, topological entanglement
and quantum computation. In this paper, we show some of the pieces in this
puzzle and how they fit together. In no way do we claim to have assembled the entire 
puzzle! That is a challenge for futher work. In order to introduce our problems, and explain what we 
have done with them, the next few paragraphs will give capsule summaries of each of the major points
of view taken in this study. We then describe in more detail what is contained in each separate 
section of the paper. The paper itself strives to be self-contained, and to describe carefully the
issues involved, particularly with topological structures that may be unfamiliar to a 
physics audience.
\bigbreak

Quantum computing can be regarded as a study of the structure of the preparation,
evolution and measurement of quantum systems. In the quantum computation model, an 
evolution is a composition of unitary transformations (finite dimensional over the complex numbers).
The unitary transformations are applied to an initial state vector that has been prepared for this process. 
Measurements are projections to elements of an orthonormal basis of the space upon which the 
evolution is applied. The result of measuring a state $| \psi \rangle,$ written in the given
basis, is probabilistic. The probability of obtaining a given basis element from the measurement is equal to the 
absolute square of the coefficient of that basis element in the state being measured. 
\bigbreak

It is remarkable that the above lines constitute an essential summary of quantum theory. 
All applications of quantum theory involve filling in details of unitary evolutions
and specifics of preparations and measurements. 
\bigbreak

One hopes to build powerful quantum computers. Such hopes would be realized if 
there were reliable ways to implement predetermined patterns of unitary evolution and measurement.
In the course of trying to understand the potential for quantum computing, it became apparent 
that arbitrary finite dimensional unitary transformations can be built from a relatively small set of 
primitives. A standard set of primitives consists in all two-dimensional unitary 
transformations, together with a choice of one sufficiently robust four-dimensional transformation
such as the $CNOT$ gate discussed in the first section of this paper. One says that 
$CNOT,$ together with single qubit gates (two dimensional unitary transformations) is {\em universal}
for quantum computation.
\bigbreak

Probability in quantum mechanics acts quite differently than classical probability. Entangled
quantum states embody this difference. An example of an entangled state is the two-qubit state
$|\psi \rangle = (|00 \rangle + |11 \rangle)/\sqrt{2}.$ This state is not decomposable as a tensor
product of single-qubit states, and a measurement in one of its tensor factors will determine the 
outcome in the other factor. Implicit in entanglement is the phenomenon of quantum non-locality:
physical access to the measurement of one tensor factor or the other may be separated by an abitrary
spatial interval. The result of a measurement can have the appearance of instantaneous determination
across an arbitrary distance.
\bigbreak

Entanglement and quantum computing are related in a myriad of ways, not the least of which is the
fact that one can replace the $CNOT$ gate by another gate $R$ and maintain universality (as described
above) just so long as $R$ can entangle quantum states. That is, $R$ can be applied
to some unentangled state to produce an entangled state. It is of interest to examine other sets of universal primitives
that are obtained by replacing $CNOT$ by such an $R.$
\bigbreak

Contemplating the inherent non-locality of entangled states, it is natural to ask whether there are 
relationships between topological entanglement and quantum entanglement. 
Topology studies global relationships in spaces, and how one space can be placed within another,
such as knotting and linking of curves in three-dimensional space. One way to 
study topological entanglement and quantum entanglement is to try making direct correspondences
between patterns of topological linking and entangled quantum states. One approach of this kind was
initiated by Aravind as we discuss in section 8 of this paper and also in \cite{TEQE,Spie}. A deeper method
(we believe) is to consider unitary gates $R$ that are both universal for quantum computation 
and are also solutions to the condition for topological braiding. Such matrices $R$ are 
unitary solutions to the Yang-Baxter equation, as explained in section 2. 
We are then in a position to compare the 
topological and quantum properties of these transformations. In this way, we can explore the
apparently complex relationship among topological entanglement, quantum entanglement, and quantum 
computational universality. It is this exploration that is the theme of this paper.
\bigbreak

In this paper, we prove that certain solutions to the Yang-Baxter equation together with 
local unitary two dimensional operators form a universal set of quantum gates. In the first version of this result, we generate $CNOT$
using a solution to the algebraic Yang-Baxter equation. In the second version, we generate $CNOT$ using versions of the
braiding Yang-Baxter equation.  Results of this kind follow from general results of the Brylinskis \cite{BB} about universal quantum
gates. Here, we give explicit proofs by expressing the $CNOT$ gate in terms of solutions to the Yang-Baxter equation (and local unitary
transformations).
\bigbreak

Section 2 of the paper defines the Yang-Baxter equation, gives unitary examples and proves the results about universal gates. We
regard these results as a significant elementary step in relating quantum topology and quantum computing. The results say that quantum
computing can be framed in the context of quantum topology. They also say that quantum computing can be framed in those statistical
mechanics contexts where the solutions to the Yang-Baxter equation are natural structures.
\bigbreak

Certainly the Yang-Baxter Equation is a natural structure in thinking about the topology of braids, knots, and links. In section 3,
we formalize an extension of the Artin Braid group that can accomodate local operators so that this extended braid group can represent
any unitary transformantion. Section 3 shows how to use solutions to the Yang-
Baxter equation to obtain such representations. The section ends with a discussion of the role of the algebraic Yang-Baxter equation in
configuring quantum circuit diagrams. 
\bigbreak

In section 4, we work out details of the invariant of knots and links that is associated with the
universal gate $R,$ and give a number of examples. In particular, we show that this invariant measures the linking of the Borromean Rings
and the Whitehead Link, both examples of links with zero linking numbers.
\bigbreak 

In section 5,  we indicate how to formulate a quantum computation of a quantum link invariant in terms of a
preparation, a unitary evolution and a measurement. We include in this context a process that quantum computes the absolute value of the
trace of an arbitrary unitary transformation. These ideas are applied in Section 7 first  to a unitary representation of the three strand
braid group that will produce a good chunk of the Jones polynomial for three strand braids when configured as a quantum computer, and
then to the invariant discussed in section 4.
\bigbreak

Entanglement is an integral ingredient in certain
communications procedures such as the teleportation of quantum states.  
In section 6 we digress on the structure of teleportation, using the ideas prsented in the previous section for obtaining the trace of a
unitary transformation. By associating a matrix $M$ to a measurement state $\langle {\cal M}|$ and using the entangled state
$|\delta \rangle$ used for preparation in the trace calculation of the previous section, we show that for unitary $M$ there is a full 
teleportation procedure for obtaining $M|\psi \rangle$ from a given state $|\psi \rangle.$ This discussion will be expanded in subsequent
papers to deal with the question of quantum computation in general, and the specific problem of computing knot invariants that
are based on non-unitary solutions to the Yang-Baxter equation. The approach to teleportation given here is inherently topological
(in the diagrammatic sense) and we shall take up its applications in subsequent papers \cite{CKL}.
\bigbreak

In section 8 we discuss the relationship between topological entanglement and quantum entanglement. We recall an invariant of links
associated with the solution to the Yang-Baxter equation used for Theorem 1. This solution, $R',$ makes an invariant
that detects linking numbers of two-component links exactly when $R'$ is capable of entangling quantum states. Examples like this, 
and invariants like the one constructed via the matrix $R,$ indicate relationships between topological entanglement
and quantum entanglement. Other examples, such as the braid group representation representing the Jones polynomial of Section 4, do not
exhibit such behaviour. The question remains open. In this section
we give an example that effectively destroys the hope of continuing an analogy of Aravind that would identify the cutting
of a link component with an observation of a state. Aravind himself showed that his notion was not invariant under basis change. 
We  point out that it is easy to build states whose entanglement or lack of it after an observation is a matter of probability obtained
from a probability amplitude. Since linking of classical links is not a matter of probability, this destroys the possibility of a direct
relationship between classical linking and quantum entanglement. Of course, there may be more subtle avenues. We are in the process of
working on such ideas. 
\bigbreak

Section 9 is a capsule summary of topological quantum field theory from the point of view of anyonic models for quantum computation.
We have included this section to indicate how braiding gates fit into a wider context.
In section 10, we carry on a philosophical discussion about the relationship of quantum and topological entanglement, speculating
that a spin network pregeometry of the right kind could enlighten us in our quest.
\bigbreak

\noindent {\bf Acknowledgement.} Most of this effort was sponsored by the Defense
Advanced Research Projects Agency (DARPA) and Air Force Research Laboratory, Air
Force Materiel Command, USAF, under agreement F30602-01-2-05022. Some of this
effort was also sponsored by the National Institute for Standards and Technology
(NIST). The U.S. Government is authorized to reproduce and distribute reprints
for Government purposes notwithstanding any copyright annotations thereon. The
views and conclusions contained herein are those of the authors and should not be
interpreted as necessarily representing the official policies or endorsements,
either expressed or implied, of the Defense Advanced Research Projects Agency,
the Air Force Research Laboratory, or the U.S. Government. (Copyright 2004.) 
It gives the first author great pleasure to acknowledge support from NSF Grant DMS-0245588,
and to give thanks to the
University of Waterloo and the Perimeter Institute in Waterloo, Canada for their hospitality during
the preparation of this research. It gives both authors pleasure  to thank Michael Nielson for very useful comments on an
early version of this paper. We thank Ferando Souza, Hilary Carteret and Niel de Beaudrap for helpful conversations.
\bigbreak

\section{Braiding Operators and Universal Gates}
We shall assume that the reader is familiar with the notions of knots, links and braids in
Euclidean three dimensional space. Recall that a knot is an embedding of a circle, taken up to 
topological equivalence, and that a link is an embedding of a collection of circles, taken up to topological
equivalence. Braids form a group under concatenation, where the concatenation of two braids is obtained by attaching
the bottom strands of the first braid to the top strands of the second braid. 
\bigbreak

A class of invariants of knots and links called quantum invariants can be constructed by using representations of 
the Artin braid group, and more specifically by using solutions to the Yang-Baxter Equation \cite{BA}, first discovered in 
relation to $1+1$ dimensional quantum field theory, and $2$ dimensional statistical mechanics. 
Braiding operators feature in constructing representations of the Artin Braid Group, and in the construction of
these invariants of knots
and links.
\bigbreak

A key concept in the construction of quantum link invariants is
the association of a Yang-Baxter operator $R$ to each elementary crossing in a
link diagram. The operator $R$ is a linear mapping  
$$R\colon \ V\otimes V \longrightarrow V\otimes V$$ 
defined on the  $2$-fold tensor product of a vector space $V,$ generalizing the permutation of the factors
(i.e., generalizing a swap gate when $V$ represents one qubit). Such transformations are not 
necessarily unitary in 
topological applications. It is a motivation for our research to understand
when they can be replaced by unitary transformations 
for the purpose of quantum 
computing. Such unitary $R$-matrices can be used to 
make unitary representations of the Artin Braid group.
\bigbreak

A solution to the Yang-Baxter equation, as described in the last 
paragraph is a matrix $R,$ regarded as a mapping of a
two-fold tensor product of a vector space
$V \otimes V$ to itself that satisfies the equation 

$$(R \otimes I)(I \otimes R)(R \otimes I) = 
(I \otimes R)(R \otimes I)(I \otimes R).$$ From the point of view of topology, the matrix $R$ 
is regarded as representing an elementary bit of braiding 
represented by one string
crossing over another. In Figure 1 below, we have illustrated 
the braiding identity that corresponds to the Yang-Baxter equation.
Each braiding picture with its three input lines (below) 
and output lines (above) corresponds to a mapping of the three fold
tensor product of the vector space $V$ to itself, as required 
by the algebraic equation quoted above. The pattern of placement of the 
crossings in the diagram corresponds to the factors 
$R \otimes I$ and $I \otimes R.$ This crucial 
topological move has an algebraic
expression in terms of such a matrix $R.$ Our main approach to 
relate topology, quantum computing, and quantum entanglement is through the 
use of the Yang-Baxter equation. In order to accomplish this aim, 
{\em we need to study solutions of the Yang-Baxter equation that are unitary.}
Then the $R$ matrix can be seen {\em either} as a braiding matrix 
{\em or} as a quantum gate in a quantum computer.

$$ \picill3inby1.5in(YangBaxterEquation) $$

\begin{center}
{\bf Figure 1  The Yang-Baxter Equation - $(R \otimes I)(I \otimes R)(R \otimes I) = 
(I \otimes R)(R \otimes I)(I \otimes R).$}
\end{center}

$$ \picill5inby4.5in(BraidGenerators) $$

\begin{center}
{\bf Figure 2 - Braid Generators and Relations }
\end{center}

The problem of finding solutions to the Yang-Baxter equation 
that are unitary turns out to be surprisingly difficult.
Dye \cite{Dye} has classified all such matrices of 
size $4 \times 4.$  A rough summary of her classification is that all $4 \times 4$ unitary 
solutions to the Yang-Baxter equation are similar to one of the following types of matrix:

\[R = \left( \begin{array}{cccc}
1/\sqrt{2} & 0 & 0  & 1/\sqrt{2}\\
0  & 1/\sqrt{2} & -1/\sqrt{2} & 0\\
0 & 1/\sqrt{2} & 1/\sqrt{2} & 0 \\
-1/\sqrt{2} & 0 & 0 & 1/\sqrt{2}\\
\end{array} \right)\] 
\bigbreak

\[R' = \left( \begin{array}{cccc}
a & 0 & 0  & 0\\
0  & 0 & b & 0\\
0 & c & 0 & 0 \\
0 & 0 & 0 & d\\
\end{array} \right)\]

\[R'' = \left( \begin{array}{cccc}
0 & 0 & 0  & a\\
0  & b & 0 & 0\\
0 & 0 & c & 0 \\
d & 0 & 0 & 0\\
\end{array} \right)\] where $a$,$b$,$c$,$d$ are unit complex numbers.

For the purpose of quantum computing, one should regard each matrix as acting on the stamdard basis
$\{ |00\rangle, |01\rangle, |10\rangle, |11\rangle \}$ of $H = V \otimes V,$ where $V$ is a two-dimensional complex vector
space. Then, for example we have
$$R|00\rangle =(1/\sqrt{2})|00\rangle-(1/\sqrt{2})|11\rangle,$$
$$R|01\rangle=(1/\sqrt{2})|01\rangle+(1/\sqrt{2})|10\rangle,$$
$$R|10\rangle=-(1/\sqrt{2})|01\rangle+(1/\sqrt{2})|10\rangle,$$
$$R|11\rangle=(1/\sqrt{2})|00\rangle+(1/\sqrt{2})|11\rangle.$$
The reader should note that $R$ is the familiar change-of-basis matrix from the standard basis to the Bell basis of entangled states.
\bigbreak

\noindent In the case of $R',$ we have
$$R'|00\rangle = a|00\rangle, R'|01\rangle= c|10\rangle,$$ 
$$R'|10\rangle = b|01\rangle, R'|11\rangle = d|11\rangle.$$ Note that $R'$ can be regarded as a diagonal
phase gate $P$, composed with a swap gate $S.$

\[P = \left( \begin{array}{cccc}
a & 0 & 0  & 0\\
0  & b & 0 & 0\\
0 & 0 & c & 0 \\
0 & 0 & 0 & d\\
\end{array} \right)\]

\[S = \left( \begin{array}{cccc}
1 & 0 & 0  & 0\\
0  & 0 & 1 & 0\\
0 & 1 & 0 & 0 \\
0 & 0 & 0 & 1\\
\end{array} \right)\] Compositions of solutions of the (Braiding) Yang-Baxter Equation with the swap gate $S$ are
called {\em solutions to the algebraic Yang-Baxter equation}. Thus the diagonal matrix $P$ is a solution to
the algebraic Yang-Baxter equation. 
\bigbreak

\subsection{Universal Gates}
A {\em two-qubit gate} $G$ is a unitary  linear mapping $G:V \otimes V \longrightarrow V$ where $V$ is a two complex dimensional
vector space. We say that the gate $G$ is {\em universal for quantum computation} (or just {\em universal}) if $G$ together with 
local unitary transformations (unitary transformations from $V$ to $V$) generates all unitary transformations of the complex vector
space of dimension $2^{n}$ to itself. It is well-known \cite{nielsen} that $CNOT$ is a universal gate. 
\bigbreak

\noindent A gate $G$, as above, is said to be {\em entangling} if there is a vector  
$$| \alpha \beta \rangle = | \alpha \rangle \otimes | \beta \rangle \in V \otimes V$$ such that 
$G | \alpha \beta \rangle$ is not decomposable as a tensor product of two qubits. Under these circumstances, one says that 
$G | \alpha \beta \rangle$ is {\em entangled}.
\bigbreak

\noindent In \cite{BB},  the Brylinskis
give a general criterion of $G$ to be universal. They prove that {\em a two-qubit gate $G$ is universal if and only if it is
entangling.} 
\bigbreak

\noindent The reader will also be interested in the paper \cite{N} and the url $$http://www.physics.uq.edu.au/gqc/,$$ wherein the
practical algorithm in \cite{N}, for expressing entangling gates in terms of CNOT and local transformations, is implemented online.
\bigbreak

\noindent It follows at once from the Brylinski Theorem that the matrices $R$, $R',$ and $R''$ are universal gates, except for certain
specific  choices of parameters in $R'$ and $R''.$ In a sequel to this paper \cite{DKLS} we will give a complete catalogue of
universality for two-qubit gates that are solutions to the Yang-Baxter equation. In this paper, we shall concentrate on specific examples
and their properties.
\bigbreak

\noindent {\bf Remark.} A two-qubit pure state $$|\phi \rangle = a|00 \rangle + b|01 \rangle + c|10 \rangle + d|11 \rangle$$
is entangled exactly when $(ad-bc) \ne 0.$ It is easy to use this fact to check when a specific matrix is, or is not, entangling.
\bigbreak

\noindent {\bf Theorem 0.} Let $D$ denote the phase gate shown below. $D$ is a solution to the algebraic Yang-Baxter equation
(see the earlier discussion in this section). Then $D$ is a universal gate.

\[D = \left( \begin{array}{cccc}
1 & 0 & 0  & 0\\
0 & 1 & 0 & 0\\
0 & 0 & 1 & 0 \\
0 & 0 & 0& -1\\
\end{array} \right)\] 

\noindent {\bf Proof.} It follows at once from the Brylinski Theorem that $D$ is universal. For a more specific proof,
note that $CNOT = QDQ^{-1},$ where
$Q=H\otimes I$, $H$ is the
$2 \times 2$ Hadamard matrix. The conclusion
then follows at once from this identity and the discussion above. 
We illustrate the matrices involved in this
proof below:
\bigbreak

\[H = (1/\sqrt{2})\left( \begin{array}{cc}
1 & 1 \\
1  & -1 \\
\end{array} \right)\]

\[Q = (1/\sqrt{2})\left( \begin{array}{cccc}
1 & 1 & 0  & 0\\
1  & -1 & 0 & 0\\
0 & 0 & 1 & 1 \\
0 & 0 & 1 & -1\\
\end{array} \right)\]

\[D = \left( \begin{array}{cccc}
1 & 0 & 0  & 0\\
0 & 1 & 0 & 0\\
0 & 0 & 1 & 0 \\
0 & 0 & 0& -1\\
\end{array} \right)\]

\[QDQ^{-1} = QDQ = \left( \begin{array}{cccc}
1 & 0 & 0  & 0\\
0 & 1 & 0 & 0\\
0 & 0 & 0 & 1 \\
0 & 0 & 1 & 0\\
\end{array} \right) = CNOT\]  This completes the proof of the Theorem. $\hfill \Box $
\bigbreak

\noindent {\bf Remark.} We thank Martin Roetteles \cite{Martin} for pointing out the specific factorization of $CNOT$ used in this
proof. \smallbreak

\noindent {\bf Theorem1.} The matrix solutions $R'$ and $R''$ to the Yang-Baxter equation, described above, are universal gates
exactly when $ad-bc \ne 0$ for their internal parameters $a,b,c,d.$ In particular, 
let $R_{0}$ denote the solution $R'$ (above) to the Yang-Baxter equation with 
$a=b=c=1, d=-1.$

\[R_0 = \left( \begin{array}{cccc}
1 & 0 & 0  & 0\\
0  & 0 & 1 & 0\\
0 & 1 & 0 & 0 \\
0 & 0 & 0 & -1\\
\end{array} \right)\]
Then $R_{0}$ is a universal gate.
\bigbreak 

\noindent {\bf Proof.} The first part follows at once from the Brylinski Theorem. 
In fact, letting $H$ be the Hadamard matrix as before, and 
\[\sigma = \left( \begin{array}{cc}
1/\sqrt{2} & i/\sqrt{2} \\
i/\sqrt{2}  & 1/\sqrt{2} \\
\end{array} \right), \, \lambda = \left( \begin{array}{cc}
1/\sqrt{2} & 1/\sqrt{2} \\
i/\sqrt{2}  & -i/\sqrt{2} \\
\end{array} \right)\]
\[\mu = \left( \begin{array}{cc}
(1-i)/2 & (1+i)/2 \\
(1-i)/2  &(-1-i)/2 \\
\end{array} \right).\]
Then $$CNOT = (\lambda \otimes \mu)(R_{0}(I \otimes \sigma)R_{0})(H \otimes H).$$
This gives an 
explicit expression for $CNOT$ in terms of $R_{0}$ and local unitary transformations
(for which we thank Ben Reichardt in response to an early version of the present paper).  
$\hfill \Box $
\bigbreak

\noindent {\bf Remark.} Let $SWAP$ denote the Yang-Baxter Solution $R'$ with $a=b=c=d=1.$

\[SWAP = \left( \begin{array}{cccc}
1 & 0 & 0  & 0\\
0  & 0 & 1 & 0\\
0 & 1 & 0 & 0 \\
0 & 0 & 0 & 1\\
\end{array} \right)\]
$SWAP$ is the standard swap gate. Note that $SWAP$ is not a universal gate. This also follows from the Brylinski Theorem, 
since $SWAP$ is not entangling. Note also that $R_{0}$ is the composition of the phase gate $D$ with this swap gate.
\bigbreak

\noindent {\bf Theorem2.} Let
\[R = \left( \begin{array}{cccc}
1/\sqrt{2} & 0 & 0  & 1/\sqrt{2}\\
0  & 1/\sqrt{2} & -1/\sqrt{2} & 0\\
0 & 1/\sqrt{2} & 1/\sqrt{2} & 0 \\
-1/\sqrt{2} & 0 & 0 & 1/\sqrt{2}\\
\end{array} \right)\] be the unitary solution to the Yang-Baxter equation discussed above. 
Then $R$ is a universal gate. The proof below gives a specific expression for $CNOT$ in terms of $R.$
\bigbreak

\noindent {\bf Proof.} This result follows at once from the Brylinksi Theorem, since $R$ is highly entangling.
For a direct computational proof, it suffices to show that $CNOT$ can be generated from $R$ and local unitary transformations.
Let
\[\alpha = \left( \begin{array}{cc}
1/\sqrt{2} & 1/\sqrt{2}\\
1/\sqrt{2} & -1/\sqrt{2}\\
\end{array} \right)\]

\[\beta = \left( \begin{array}{cc}
-1/\sqrt{2} & 1/\sqrt{2}\\
i/\sqrt{2} & i/\sqrt{2}\\
\end{array} \right)\]

\[\gamma = \left( \begin{array}{cc}
1/\sqrt{2} & i/\sqrt{2}\\
1/\sqrt{2} & -i/\sqrt{2}\\
\end{array} \right)\]

\[\delta = \left( \begin{array}{cc}
-1 & 0\\
0 & -i\\
\end{array} \right)\]  Let $M= \alpha \otimes \beta$ and $N= \gamma \otimes \delta.$ Then it is straightforward to verify that
$$CNOT = MRN.$$ This completes the proof. $\hfill \Box $
\bigbreak

\noindent {\bf Remark.}  
We take both Theorems $1$ and $2$ as
suggestive of fruitful interactions between quantum topology and quantum  computing. It is worth comparing these Theorems
with the results in \cite{FLZ}, a comparison that we shall leave to  a future paper.
\bigbreak

\noindent {\bf Remark.} We thank Stephen Bullock for his help in obtaining this result. On showing him the Yang-Baxter solution
$R$ used in the above proof, he showed us the paper ($\cite{Bullock}$ by V. V. Shende, S. S. Bullock and I. L. Markov) in which  he
and his coauthors give a criterion for determining if a $4 \times 4$  unitary matrix can be generated by local unitary transformations
and a single
$CNOT.$ We then calculated that criterion and  found that $R$ passes the test. Bullock then showed us how to apply their theory to obtain
the specific transformations needed in this case. Thus the above result is a direct application of their paper. The criterion also
shows that the solutions of type
$R^{'}$ and $R^{''}$ listed above require two applications of $CNOT.$ We will discuss their structure elsewhere; but for the record,
it is of interest here to record the Shende,Bullock,Markov criterion.
\bigbreak

\noindent {\bf Theorem \cite{Bullock}.}  We shall say that a matrix can be {\em simulated} using $k$ $CNOT$ gates if it can
be expressed by that number $k$ of $CNOT$ gates plus local unitary transformations. Let $E$ be the following matrix:
\[E = \left( \begin{array}{cccc}
0 & 0 & 0  & 1\\
0  & 0 & -1 & 0\\
0 & -1 & 0 & 0 \\
1 & 0 & 0 & 0\\
\end{array} \right)\]
Let $U$ be a matrix in $SU(4).$ Let $\gamma(U)$ be defined by the formula
$$\gamma(U) = UEUE.$$ Let $tr(M)$ denote the trace of a square matrix $M.$ Then 
\begin{enumerate}
\item $U$ can be simulated using zero $CNOTS$ if and only if $\gamma(U) = I,$ where $I$ denotes the identity matrix.
\item $U$ can be simulated using one $CNOT$ gate if and only if $tr[\gamma(U)]=0$ and $\gamma(U)^{2} = -I.$
\item $U$ can be simutated using two $CNOT$ gates if and only if $tr[\gamma(U)]$ is real.
\end{enumerate}
$\hfill \Box $
\bigbreak

\noindent Note that in applying this criterion, the matrix in question must be in the special unitary group. We leave it to the reader
to show that matrices of type $R'$ and $R''$ require two $CNOTS,$ and that the matrix $R$ is picked by this criterion
to require only one $CNOT,$ just as we have shown explicitly above. Note that since $R^{8}$ is the 
identity, we have $R^{-1} = R^{7},$ as well as the fact that $R^{-1}$ can be expressed in terms of local transformations and a single
application of $CNOT.$
\bigbreak

\section{Generalizing and Representing the Artin Braid Group} 
Let $B_{n}$ denote the Artin braid group on $n$ strands \cite{Sossinsky}.
We recall here that $B_{n}$ is generated by elementary braids $\{ s_{1}, \cdots ,s_{n-1} \}$
with relations 

\begin{enumerate}
\item $s_{i} s_{j} = s_{j} s_{i}$ for $|i-j| > 1$, 
\item $s_{i} s_{i+1} s_{i} = s_{i+1} s_{i} s_{i+1}$ for $i= 1, \cdots n-2.$
\end{enumerate}

\noindent See Figure 2 for an illustration of the elementary braids and their relations. Note that the braid group has a diagrammatic
topological interpretation, where a braid is an intertwining of strands that lead from one set of $n$ points to another set of $n$ points.
The braid generators $s_i$ are represented by diagrams where the $i$-th and $(i + 1)$-th strands wind around one another by a single 
half-twist (the sense of this turn is shown in Figure 2) and all other strands drop straight to the bottom. Braids are diagrammed
vertically as in Figure 2, and the products are taken in order from top to bottom. The product of two braid diagrams is accomplished by
adjoining the top strands of one braid to the bottom strands of the other braid. 
\bigbreak 

In Figure 2 we have restricted the illustration to the
four-stranded braid group $B_4.$ In that figure the three braid generators of $B_4$ are shown, and then the inverse of the
first generator is drawn. Following this, one sees the identities $s_{1} s_{1}^{-1} = 1$ 
(where the identity element in $B_{4}$ consists in  four vertical strands), 
$s_{1} s_{2} s_{1} = s_{2} s_{1}s_{2},$ and finally
$s_1 s_3 = s_3 s_1.$ 
With this interpretation, it is apparent from Figures 1 and 2 that the second braiding relation (above) is formally the same as the
Yang-Baxter equation.
\bigbreak

In fact, if $V$ denotes the basic vector space (the space for one qubit in our context), and $R$ is an invertible solution
to the Yang-Baxter equation
as described in section 2, then we obtain a representation of the $n$ strand braid group into the vector
space of automorphisms of the $n$-th tensor power of $V:$
$$rep_{n}:B_n \longrightarrow Aut(V^{\otimes n})$$
by defining $$rep_{n}(s_{i}) = I^{\otimes i-1} \otimes R \otimes I^{\otimes n-i-1}$$ 
where $I$ denotes the identity mapping on $V.$ Note that since $R$ is a unitary matrix, it follows that this representation is unitary.
We shall call this the {\em standard} method for making a representation of the braid group from an invertible solution to the 
Yang-Baxter equation. Note that $rep_{n}(s_{i})$ is supported by $R$ on the $i$-th and $(i + 1)$-th tensor factors and is the
identity  mapping on the other factors.
\bigbreak

We now wish to generalize the classical Artin braid group to a larger group whose representations can include compositions of
the elements $rep_{n}(s_{i})$ constructed in the last paragraph with local unitary transfomations. A diagrammatic example of such 
a composition is given in Figure 3, where we illustrate (as in Theorem $2$) the expression of $CNOT$ in terms of $R$ and local unitary
transfomations. In this diagram, the local unitary transformations are inicated by nodes on single braiding lines. This corresponds
to the fact that the local unitary operations act on single tensor factors. Thus we shall generalize the braid group so that the new
group generators are represented on single tensor factors while the elementary braids are represented on two essential tensor factors.
This is formalized in the next paragraph.
\bigbreak 

Let $G$ be any group, and let $G^{\otimes n}$ denote the $n$-fold tensor product of $G$ with itself, where by this tensor
product we mean the group whose elements are of the form $g_1 \otimes g_2 \otimes \cdots \otimes g_n,$  with $g_i \in G,$ satisfying
$$(g_1 \otimes g_2 \otimes \cdots \otimes g_n)(g_{1}' \otimes g_{2}' \otimes \cdots \otimes g_{n}') = g_1 g{_1}' \otimes g_2 g{_2}'
\otimes \cdots \otimes g_n g{_n}'.$$  We articulate $G^{\otimes n}$ in tensor language,
because we wish to consider representations of the group where individual members of $G$ go to matrices and the elements $g_1 \otimes g_2
\otimes \cdots \otimes g_n$ are sent to tensor products of these matrices. Let $h_{i}(g) = e \otimes e \cdots \otimes g \otimes e \cdots
\otimes e,$ where $e$ is the identity element in $G,$ and the element $g$ is in the $i$-th place in this tensor product string.
Then $$g_1 \otimes g_2 \otimes \cdots \otimes g_n = h_{1}(g_1) \cdots h_{n}(g_n),$$ and $G^{\otimes n}$ is generated by
the elements $h_{i}(g)$ where $i$ ranges from $1$ to $n$ and $g$ ranges over elements of $G.$ Note that for $i \ne j$, and 
for any $g, g' \in G,$ the elements
$h_{i}(g)$ and $h_{j}(g')$ commmute.  Note that $h_i$ is an isomorphism of $G$ to the subgroup $h_{i}(G)$ of
$G^{\otimes n}.$
\bigbreak

We define an extension $GB_n$ of the braid group $B_{n}$ by the group $G^{\otimes n}$ as follows: $GB_n$ is freely generated
by $G^{\otimes n}$ and $B_n$ modulo the relations $$h_{i}(g)s_j = s_{j} h_{i}(g)$$ for all $g$ in $G$ and all choices of
$i$ and $j$ such that $i < j$ or $i > j+1.$
\bigbreak

Just as there is a diagrammatic intepretation of the braid group $B_{n}$  in terms of strings that entangle one another, there is
a diagrammatic intepretation of $GB_{n}.$  Think of a braid diagram and suppose that on the lines of
the diagram there is a collection of labelled dots, with each dot labeled by an element of the group $G.$ Let it be given that if two
dots occur consecutively on one of the strings of the braid diagram, then they can be replaced by a dot labeled by the product of these
two elements. We make no assumptions about moving dots past elementary braiding elements (which have the appearance of one string passing
over or under the other). It is easy to see that this diagrammatic description also defines a group extending the braid group, and that
this diagrammatic group is isomorphic to $GB_n.$ 
\bigbreak

We apply this description of $GB_n$ by taking $G=U(2)$, the $2 \times 2$ unitary matrices, viewing them as local unitary transformations
for quantum computing. Then we let $UB_n$ denote $U(2)B_n$ and take the representation of $UB_n$ to $Aut(V^{\otimes n})$ that is obtained
by the mapping $$\Gamma: UB_n \longrightarrow Aut(V^{\otimes n})$$ defined by  $\Gamma(h_{i}(g)) = h_{i}(g)$ for $g$ in $U(2)$ and 
$\Gamma(s_{i}) = I^{\otimes i-1} \otimes R \otimes I^{\otimes n-i-1}$ where $R$ is the Yang-Baxter solution discussed in 
Theorem $2.$ 
\bigbreak

Recall that a quantum computer is an $n$-qubit unitary transformation $U$ coupled with rules/apparatus
for preparation and measurement of quantum states to which this transformation is applied. We then conclude from Theorem $2$ that 
\bigbreak

\noindent {\bf Theorem 3.}   Any quantum computer  has its basic unitary
transformation $U$ in the image of $\Gamma.$ 
\bigbreak

$$ \picill3inby3in(ExtraElements) $$

\begin{center}
{\bf Figure 3 - $CNOT = (A \otimes B)R(C \otimes D)$ }
\end{center}

\noindent {\bf Remark.} This theorem means that, in principle, one can draw a circuit diagram for a quantum computer that is written
in the language of the extended braid group $UB_{n}.$ In particular this means that braiding relations will apply for sectors of the 
circuitry that are not encumbered by local unitary transformations. Typically, there will be many such local unitary transformations.
We will investigate this braid algebraic structure of quantum computers in a sequel to this paper. 
A key illustration of Theorem $3$ is the diagrammatic interpretation of Theorem $2.$ This is shown in Figure 3 where we have written in
diagrams an equation of the form  $$CNOT = (A \otimes B)R(C \otimes D).$$ Here $A$, $B$, $C$ and $D$ represent the local unitary 
matrices that are used in Theorem $2$ (with different names) to express $CNOT$ in terms of $R.$
\bigbreak 

In general, a unitary transformation can
be written as an extended braiding diagram, with appearances of $R$ occupying two adjacent strands, and of local transformations occupying
single strands. Note that since $R^{8}$ is the identity, Theorem 3 actually says that a unitary transformation can be built via a
representation of the extension of a quotient $B_{n}'$ of the braid group $B_{n},$ where each braid generator $s_{i}'$ in $B'_{n}$ has
order eight. It is worth investigating the algebraic structure of $B'_{n}.$ This is a topic for further research.
\bigbreak

\subsection{The Algebraic Yang-Baxter Equation}
If $R$ denotes a solution to the Yang-Baxter equation (not neccessarily the $R$ of Theorem $2$), then we can consider the composition
$r = SR$ where $S$ is the swap gate defined in section 2. If we think of $r$ as supported on two tensor lines, and write
$r_{i \, j}$ for the same matrix, now supported on tensor lines $i$ and $j,$ (all other lines carrying the identity matrix) then we find
that the Yang-Baxter equation for $R$ is equivalent to the following equations for $r_{i \, j}.$
$$r_{i \,  i+1} \, r_{i \, i+2} \, r_{i+1 \, i+2} =  r_{i+1 \, i+2} \, r_{i \, i+2} \, r_{i \, i+1}.$$
The above equation is called the {\em Algebraic Yang-Baxter Equation.} See Figure 4 for an illustration of this relationship.
In making circuit diagrams to apply Theorem $1,$ it is useful to use the formalism of the algebraic Yang-Baxter equation since
we can then think of the phase gate $D$ as such a solution and use the above relation to relocate compositions of $D$ on diffferent
tensor lines. Given a solution to the algebraic Yang-Baxter equation, plus the swap gate $S$ we can again define a generalization of the 
Braid Group that includes local unitary transformations on the single tensor lines.  We
will leave detailed application of this point of view to a sequel to this paper.
\bigbreak

$$ \picill3inby4.5in(AlgYBE) $$

\begin{center}
{\bf Figure 4 -Algebraic Yang Baxter Equation }
\end{center}

\section{An Invariant of Knots and Links Associated with the Matrix $R$}
A well-known relationship between braids and knots and links allows the construction of invariants of knots and links
from representations of the Artin Braid Group. We give here a quick summary of these relationships and then apply them to the 
quantum universal matrix $R,$ showing that it gives rise to an interesting invariant of knots and links. The reader should note 
that this section is concerned only with the classical braid group. It does not use the extensions of the braid group that are discussed
in the previous section.
\bigbreak

At this point it is worth making a digression about the Reidemeister moves. In the 1920's Kurt Reidemeister
proved an elementary and important theorem that translated the problem of determining the topological type of a knot
or link to a problem in combinatorics. Reidemeister observed that any knot or link could be represented by a 
{\em diagram,} where a diagram is a graph in the plane with four edges locally incident to each node, and with extra
structure at each node that indicates an over-crossing of one local arc (consisting in two local edges in the graph)
with another. See Figure 5. The diagram of a classical knot or link has the appearance of a sketch of the knot; but it
is a rigorous and exact notation that represents the topological type of the knot. Reidemeister showed 
that two diagrams represent the same topological type (of knottedness or linkedness) if and only if one diagram can be
obtained from another by planar homeomorphisms coupled with a finite sequence of the {\em Reidemeister moves} \
illustrated in Figure 6. Each of the Reidemeister moves is a local change in the diagram that is applied as shown in 
this Figure.

$$ \picill3inby2in(Knot)  $$

\begin{center}
{\bf Figure 5 - A Knot Diagram}
\end{center}

$$ \picill3inby3in(Moves)  $$

\begin{center}
{\bf Figure 6 - Reidemeister Moves}
\end{center}

$$ \picill3inby3in(BraidClosure)  $$

\begin{center}
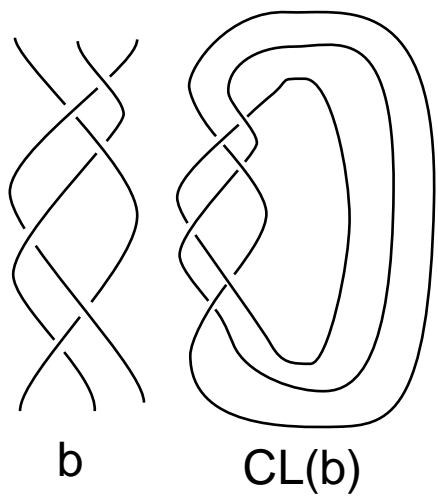
{\bf Figure 7 - Closing a Braid to form the Borromean Rings}
\end{center}

We say that two knots or links are {\em isotopic}
if one can be obtained from the other by a sequence using any of the three Reidemeister moves (plus global topological mappings of 
the diagram plane to itself).
\bigbreak

The first significant fact relating links and braids is the 
\smallbreak

\noindent {\bf Theorem of Alexander.} Every knot or link is isotopic to the closure of a braid.
\bigbreak

\noindent The closure of a braid $b$, here denoted $CL(b),$ is obtained by attaching each top strand to the corresponding bottom 
strand in the fashion shown in Figure 7. The closed braid is a weave that proceeds circularly around a given axis. There are many proofs
of Alexander's Theorem. The interested reader should consult \cite{Sossinsky}. 
\bigbreak

$$ \picill4inby3.5in(links)  $$

\begin{center}
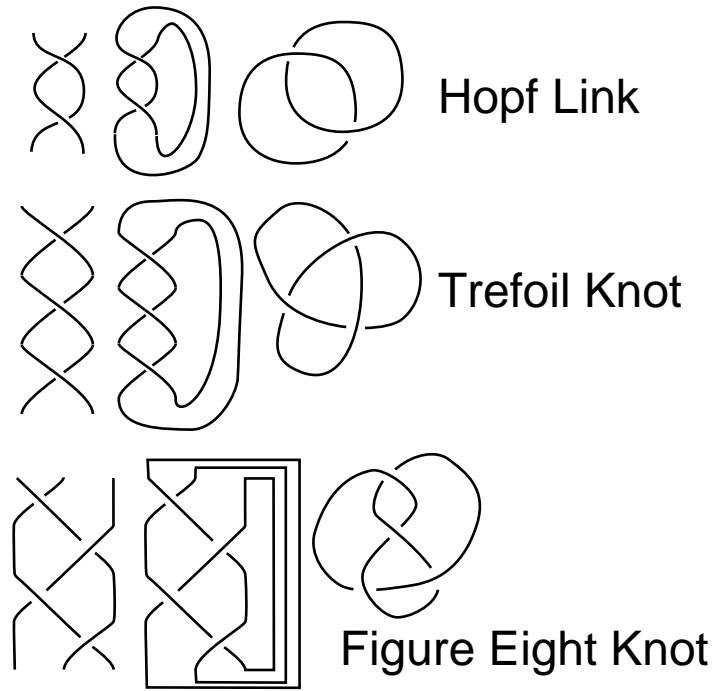
{\bf Figure 8 - Closing  Braids to Produce Hopf Link, Trefoil Knot and Figure Eight Knot}
\end{center}

Given that every knot or link can be represented by a closed braid, it is natural to wonder whether the classification of braids will
effect a classification of the topological types of all knots and links. The situation is more complicated than one might have expected.
There are many braids whose closure is isotopic to any given knot or link. Here are two basic methods for modifying a braid $b$
in $B_{n}$ so that the topological type of its closure does not change:
\begin{enumerate}
\item Let $g$ be any braid in $B_{n}.$ Then $$CL(gbg^{-1}) = CL(b)$$ where we use equality to denote isotopy of knots and links as 
described above.
\item Note that if $b$ is in $B_{n},$ then $b s_n$ is in $B_{n+1}.$ It is easy to see that $$CL(b) = CL(b s_{n})$$
and $$CL(b) = CL(b s_{n}^{-1}).$$
\end{enumerate}
\bigbreak

In light of the equvalences we have just indicated, the following two moves on braids are called the {\em Markov moves}
(after Markov who enunciated the Theorem we state below):
\begin{enumerate}
\item {\bf Markov Move 1} Replace a braid $b$ by $gbg^{-1}$ where $g$ is another braid with the same number of strands.
\item {\bf Markov Move 2} Replace a braid $b \in B_n$ by either $bs_n$ or by $bs_{n}^{-1}$ or vice versa, replace
$bs_{n}^{\pm 1}$ with $b.$
\end{enumerate}
\bigbreak

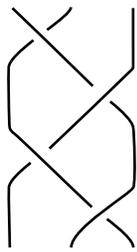
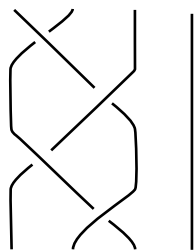
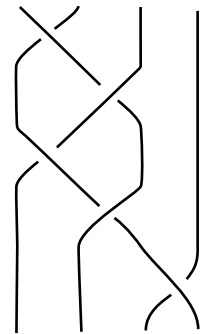
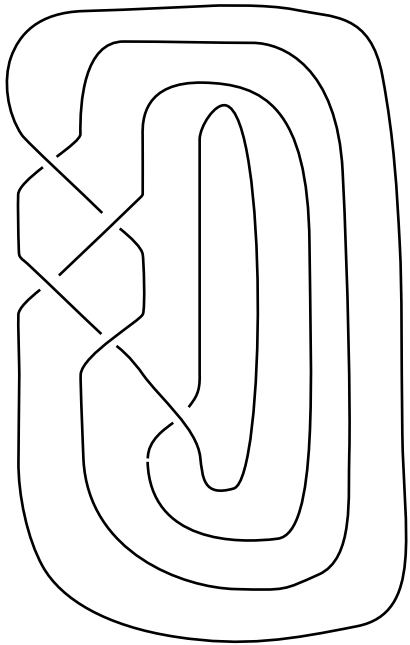
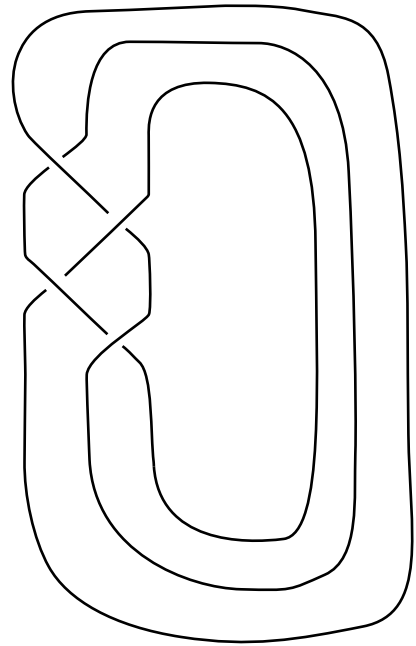
$$ \picill4inby4.5in(MarkovMove)  $$

\begin{center}
{\bf Figure 9 - Illustration of the Second Markov Move}
\end{center}

$$ \picill4inby6in(MarkovTrace)  $$

\begin{center}
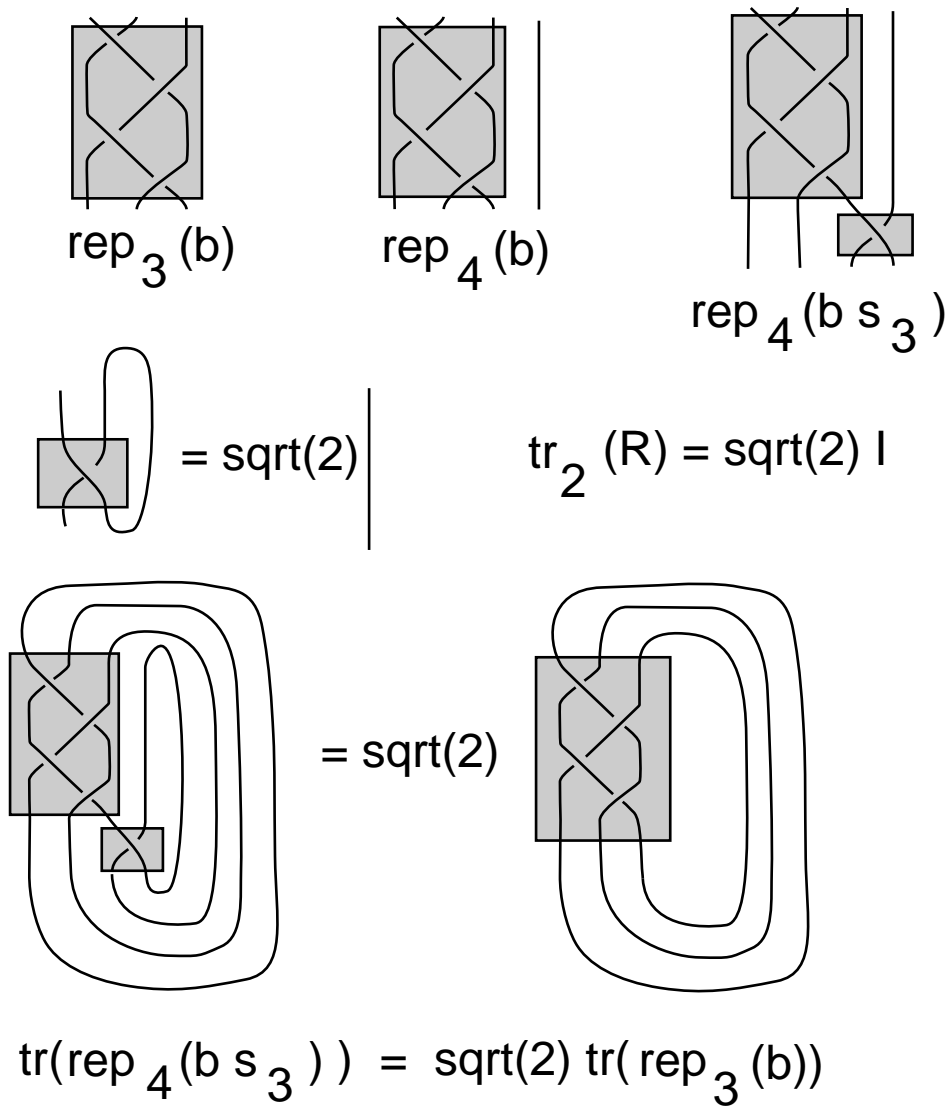
{\bf Figure 10 - Illustration of the Behaviour of the Trace on the Second Markov Move}
\end{center}
\vspace{20mm}

\noindent {\bf Markov Theorem.} Suppose that $b$ and $b'$ are two braids (of possibly different numbers of strands) with 
$CL(b) = CL(b').$ Then $b'$ can be obtained from $b$ by a series of braid equivalences coupled with applications of the 
Markov moves).
\bigbreak

\noindent {\bf Remark.} For proofs of the Markov Theorem, see \cite{Birman} and \cite{Lambropoulou}.
See Figure 9 for an illustration of the second Markov move. Notice that in making this move we promote the braid $b \in B_{n}$ to a
braid in $B_{n+1}$ by adding a right-most strand. Then we multiply by $s_{n} \in B_{n+1}.$ The closure of the resulting braid
differs by a single first Reidemeister move from the closure of $b.$
The upshot of this theorem is that it is {\em possible} for a trace function on a representation of the braid group to give rise
to topological information about the closure of the braid.  For example, suppose that we have a unitary representation of the braid
group arising from a unitary solution of the Yang-Baxter equation, as described in Section 3.  Let the representation be denoted by
$$rep_{n}: B_n \longrightarrow Aut(V^{\otimes n}).$$ Let $$\tau(b) = tr(rep_{n}(b))$$ where $tr(M)$ denotes the trace of a square matrix
$M.$ 
Then, since the trace of any linear mapping satisfies
$tr(AB) = tr(BA),$ it follows that $\tau(gbg^{-1}) = \tau(b),$ and hence $\tau$ gives the same values on braids that differ by 
Markov moves of type $1.$ We would like $\tau$ to be invariant under Markov moves of type $2,$ but this is usually too much to ask.
It is standard practice in the literature of link invariants to search for a matrix $\eta$ mapping $V$ to $V$ such that the modified
trace $TR(b) = tr(\eta^{\otimes n}rep_{n}(b))$ has a multiplicative property under the second Markov move in the sense that
$TR(b s_{n}) = \alpha TR(b)$ and  $TR(b s_{n}^{-1}) = \alpha^{-1} TR(b),$ where $\alpha$ is a invertible constant in the ring
of values for the trace. Such a function $TR$ is called a {\em Markov trace}, and one can normalize it to obtain a function that is an
invariant of isotopy of links by defining $I(b) = \alpha^{-w(b)}TR(b),$ where $w(b)$ is the sum of the signs of the crossings of the 
braid $b.$
\bigbreak

In the case of our computationally universal matrix $R,$ the bare trace $\tau(b) = tr(rep_{n}(b))$
behaves in a very simple way under the second Markov move. We find (and will show the details below) that 
$$\tau(b s_{n}) = \sqrt{2} \tau(b)$$ and 
$$\tau(b s_{n}^{-1}) = \sqrt{2} \tau(b).$$
Note that the multiplicative factor is the same for both types of second Markov move. Instead of making a normalizing factor from 
this, we can say that if two links $CL(b)$ and $CL(b')$ are isotopic, then $\tau(b)$ and $\tau(b')$ will differ by a multiplicative
factor that is some power of the square root of two. In particular, this means that if $\tau(b)$ and $\tau(b')$ have different sign, or
if one is zero and the other not zero, then we know that the closures of $b$ and $b'$ are not isotopic.
\bigbreak

\noindent Here is the matrix $R.$

\[R = \left( \begin{array}{cccc}
1/\sqrt{2} & 0 & 0  & 1/\sqrt{2}\\
0  & 1/\sqrt{2} & -1/\sqrt{2} & 0\\
0 & 1/\sqrt{2} & 1/\sqrt{2} & 0 \\
-1/\sqrt{2} & 0 & 0 & 1/\sqrt{2}\\
\end{array} \right)\]  
\bigbreak
\noindent We have
$$R|ab\rangle = R_{ab}^{00}|00\rangle + R_{ab}^{01}|01\rangle + R_{ab}^{10}|10\rangle + R_{ab}^{11}|11\rangle.$$
 
\noindent Let 
$$tr_{2}(R) = \Sigma_{k} R_{ak}^{bk}$$ be the partial trace of $R$ with respect to the second tensor factor.
Then
$$tr_{2}(R) = (1/\sqrt{2} + 1/\sqrt{2}) \,I = \sqrt{2} \, I$$  where $I$ denotes the $2 \times 2$ identity matrix and 
$$tr_{2}(R^{-1}) = \sqrt{2} \, I.$$  

\noindent Recall from the previous section that 
$$rep_{n}:B_n \longrightarrow Aut(V^{\otimes n})$$
is defined on braid generators by the equation
$$rep_{n}(s_{i}) = I^{\otimes i-1} \otimes R \otimes I^{\otimes n-i-1}$$ 
where $I$ denotes the identity mapping on $V.$
Now suppose that $b \in B_{n}.$ We want to compute $\tau(bs_{n}).$ This expression requires interpretation. When we write
$bs_n,$ we are taking $b$ in $B_n$ and regarding it as an element of $B_{n+1}$ by adding an extra right-most strand to $b.$
In general, by adding strands in this way, we have standard embeddings of $B_n$ in $B_m$ when $m \ge n.$ Working in $B_{n+1},$
we have $$rep_{n+1}(b) = rep_{n}(b) \otimes I$$ and $$rep_{n+1}(s_{n}) = I^{\otimes (n-1)} \otimes R.$$ 
Thus
$$\tau(b s_n) = tr(rep_{n+1}(bs_{n})) = tr((rep(b)\otimes I)(I^{\otimes (n-1)} \otimes R)).$$ 
From this it is easy to see that in tracing $rep_{n+1}(bs_{n}),$ the rightmost indices of the matrix $R$ (in the $n+1$ tensor factor) are
contracted directly with one another (since $b$ is supported on the first $n$ strands). Thus the partial trace is applied to the $R$ that
appears in the representation of $bs_{n}$ corresponding to $s_{n}.$ It follows from this that

$$\tau(bs_n) = tr(rep_{n}(b))tr_{2}(R) = \tau(b) \sqrt{2}.$$
Hence, for the tensor representation
built from $R$ as described in section 3, we have 
$$\tau(bs_{n}) = \sqrt{2} \tau(b)$$ and in like manner, we have 
$$\tau(bs_{n}^{-1}) = \sqrt{2} \tau(b).$$
This proves the assertions we have made about the properties of $\tau$ for this $R.$

\noindent {\bf Remark.} In Figure 10, we illustrate diagrammatically the above argument at the index level.
In this illustration, we have placed a shaded box around a braid to indicate the application of the representation of the braid
group. Thus a shaded braided box represents a matrix with upper indices correpsponding to the upper strands on the box, and lower indices
corresponding to the lower strands on the box. The simplest instance of such a matrix is a single vertical line which represents the 
identity matrix, and iconically indicates the identity of the top index with the bottom index (hence representing the identity as a 
Kronecker delta).
\bigbreak

\noindent It is a fact that shaded boxes so placed on the
braids give a correct picture of the contractions of the corresponding matrices via the convention that {\em we contract the indices
along lines that connect free index ends between diagrammatic matrices.} The figure then illustrates directly via these diagrammatic
matrices how we obtain the formula
$$tr(rep_{n+1}(b s_n)) =\sqrt{2} tr(rep_{n}(b)).$$ Note that the trace of a  diagrammatic matrix has exactly the same form as
the closure of a braid, since the connection of two open lines corresponds to the  identification and contraction over their respective
indices.

Finally, here is the same argument using matrix algebra with indices. We use the Einstein summation convention: Summation is taken over
repeated upper and lower indices. Suppose that 
$rep_{n}(b) = (M^{a,i}_{b,j})$ where $a$ and $b$ are vectors of indices for the the first $n-1$ factors of the tensor product, and 
$i$ and $j$ are individual indices with values $0$ or $1.$ Then $$rep_{n+1}(b) = (M^{a,i}_{b,j} \delta^{r}_{s})$$ where
$\delta^{r}_{s}$ is the $2 \times 2$ identity matrix. Furthermore $$rep_{n+1}(s_n) = \delta^{a}_{b} R^{t,u}_{v,w}.$$
Hence $$rep_{n+1}(b s_n) =(M^{a,i}_{b,j}R^{j,u}_{k,w})$$ from which it follows that 
$$tr(rep_{n+1}(b s_n)) = tr(M^{a,i}_{b,j}R^{j,u}_{k,w}) = tr(M^{a,i}_{b,j}R^{j,u}_{k,u})$$
where {\em in the last equality we have contracted the last indices of $R.$} Since $$R^{j,u}_{k,u} = \sqrt{2} \delta^{j}_{k}$$
it follows that 
$$tr(rep_{n+1}(b s_n)) = tr(M^{a,i}_{b,j} \sqrt{2} \delta^{j}_{k}) = \sqrt{2} tr(M^{a,i}_{b,k}) = \sqrt{2} tr(rep_{n}(b)).$$
This completes the explicit index verification of the behaviour of $\tau$ under the second Markov move.
\bigbreak

This invariant of knots and links turns out to be quite interesting. For example,
it detects the linkedness of the Borromean rings (depicted in Figures 7 and 16). It gives the following values:
\begin{enumerate}
\item $\tau(Unlink \ of \ Three \ Components) = 8 >0$
\item $\tau(Hopf \ Link) = 0$
\item $\tau(Trefoil \ Knot) = -2 \sqrt{2} < 0$
\item $\tau(Figure \ Eight \ Knot) = -4 < 0$
\item $\tau(Borromean \ Rings) = -8 < 0$
\end{enumerate}

Note that $\tau$ does not detect the difference between the trefoil knot, the figure  eight knot and the Borromean rings, but it does
show that the Hopf link is linked, that the Borromean rings are linked, and that the trefoil knot and the figure eight knot are knotted.
See Figure 8 for illustrations of these knots  and links. It remains to be seen how the 
quantum entangling properties of the matrix $R$ are related to the behaviour of this link invariant.
\bigbreak

\noindent {\bf Remark on a Skein Relation.} 
In this subsection, we point out that there is a {\em skein relation} that helps in the computation of the trace
$\tau(b)$ for a braid $b.$ A skein relation is an equation about an invariant involving local changes at the site of a single crossing
in  corresponding braid or link diagrams. The first skein relation in knot theory was discovered and utilized for the Alexander
polynomial by  John H. Conway in his remarkable paper \cite{Conway}. (Conway used an idea that was implicit in Alexander's original paper
of 1928.) The Jones polynomial and many other knot polynomial invariants satisfy such relations.
\bigbreak

The matrix $R$ satisfies the equation $$R + R^{-1} = \sqrt{2}I_{2}$$ where $I_{n}$ denotes
the $2^{n} \times 2^{n}$ identity matrix. 
We leave it to the reader to check this fact. It is also easy to check that $$R^8=I_2$$ and that all the
lower powers are non-trivial. The fact that $R$ has finite order certainly limits its power as a link or braid invariant.
For example, we have the eight-fold periodicity $$\tau(s_{i}^{n+8}) = \tau(s_{i}^{n})$$ as a direct consequence of the finite
order of 
$R.$ On the other hand the identity $R+R^{-1} = \sqrt{2}I_{2}$ can be viewed as a method for simplifying the calculations for a braid.
This implies the skein relation $$\tau(b) + \tau(b') = \sqrt{2}\tau(b'') $$ when $b$ and $b'$ are elements of the
$n$-strand braid  group that differ at a single crossing and $b''$ is the result of replacing this crossing by an identity braid. 
The crossing can be interpreted as a single instance of
$s_{i}$ for some $i,$ and we then use $rep_{n}(\alpha  s_{i} \  \beta) + rep_{n}(\alpha s_{i}^{-1} \beta) = \sqrt{2} \ rep_{n}(\alpha 
\beta).$ 
\bigbreak

\noindent {\bf Example 1.} Here is the simplest example of this  sort of computation. We work in $B_{2}$ and let $s = s_{1}:$ 
$$\tau(s s) + \tau(s s^{-1}) = \sqrt{2}\tau(s).$$  
Here we have $\tau(s s^{-1}) = \tau(I_{2}) = 4$ and $\tau(s) = \sqrt{2}\tau(I_{1}) = 2 \sqrt{2}.$ Hence
$\tau(s s) = 0,$ as we remarked earlier with $s s$ the braid representative for the Hopf link (See Figure 8). 
More generally, we have $$\tau(s^{n+1}) + \tau(s^{n-1}) = \sqrt{2} \tau(s^{n})$$ so that
$$\tau(s^{n+1}) =  \sqrt{2} \tau(s^{n}) - \tau(s^{n-1}).$$
Letting $1$ denote the identity braid in $B_{2},$ we then have
$$\tau(1) = 4$$
$$\tau(s) = 2 \sqrt{2}$$
$$\tau(s^2) = 0$$
$$\tau(s^3) = -2 \sqrt{2}$$
$$\tau(s^4) = -4$$
$$\tau(s^5) = -2 \sqrt{2}$$
$$\tau(s^6) = 0$$
$$\tau(s^7) = 2 \sqrt{2}$$
with the periodicity $$\tau(s^{n+8}) = \tau(s^{n}).$$ Since $s^7$ and $s^3$ close to knots, we see that this invariant can distinguish
these two knots from one another, but cannot tell that the closure of $s^7$ is knotted.

\noindent {\bf Example 2.} Let $b = s_{1}^{2}s_{2}^{-1}s_{1}s_{2}^{-1}.$ See Figure 11. The closure of $b$ is $W,$ a link of two 
components, with linking number equal to zero. $W$ is called the {\em Whitehead Link,} after the topologist, J. H. C. Whitehead, who
first studied its properties.  We shall check that
$\tau(b) = -4\sqrt{2},$ showing that our invariant detects the  linkedness of the Whitehead link.
\bigbreak

We use skein relation for the first appearance of $s_{2}^{-1}$ from the left on the word for $b.$ This gives
$$\tau(b) = - \tau(s_{1}^{2}s_{2}s_{1}s_{2}^{-1}) + \sqrt{2}\tau(s_{1}^{2}s_{1}s_{2}^{-1}).$$
Note that $$s_{1}^{2}s_{2}s_{1}s_{2}^{-1} = s_{1}(s_{1}s_{2}s_{1})s_{2}^{-1} =  s_{1}(s_{2}s_{1}s_{2})s_{2}^{-1} =
s_{1}s_{2}s_{1}.$$ Then $$\tau(s_{1}s_{2}s_{1}) = \tau((s_{1}s_{2})s_{1}) = \tau(s_{1}(s_{1}s_{2})) = \tau(s_{1}^{2}s_{2})
=\sqrt{2}\tau(s_{1}^{2}) = 0.$$ Hence
$$\tau(b) =  \sqrt{2}\tau(s_{1}^{2}s_{1}s_{2}^{-1}) = \sqrt{2}\tau(s_{1}^{3}s_{2}^{-1}) = \sqrt{2}^{2}\tau(s_{1}^{3}) = 2(-2\sqrt{2})
= -4\sqrt{2}.$$

$$ \picill4inby5.5in(WhiteheadLink)  $$

\begin{center}
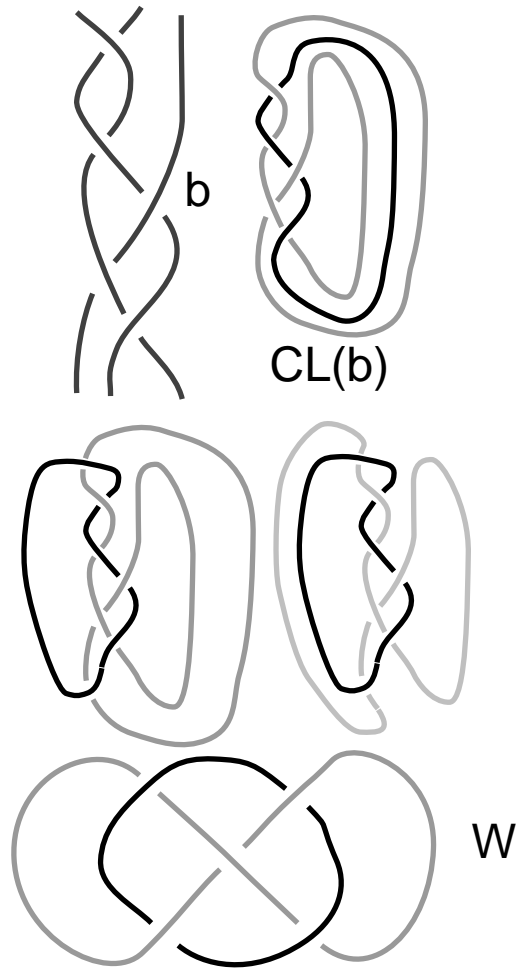
{\bf Figure 11 - Whitehead Link, $W = CL(b =s_{1}^{2}s_{2}^{-1}s_{1}s_{2}^{-1}).$}
\end{center}

\section{Quantum Computation of Knot Invariants}
 
{\em Can the invariants of knots and links such as the Jones 
polynomial be configured as quantum computers?} This is an 
important question because the algorithms to compute the Jones 
polynomial are known to be $NP$-hard, and 
so corresponding quantum algorithms may shed light on the 
relationship of this level of computational complexity with 
quantum computing (See \cite{BFLW}).  Such models can be formulated in terms of 
the Yang-Baxter equation \cite{KA87, KA89,KP, QCJP}. 
The next paragraph explains how this comes about.

In Figure 12, we indicate how topological braiding plus 
maxima (caps) and minima (cups) can be used to configure the 
diagram of a knot or link. This also can be translated into algebra 
by the association of a Yang-Baxter matrix $R$ (not necessarily the $R$ of the previous sections) to each 
crossing and other matrices
to the maxima and minima. There are models 
of very effective invariants of knots and links such as 
the Jones polynomial that can be 
put into this form \cite{QCJP}. In this way of looking at 
things, the knot diagram can be viewed as a picture,
with time as the vertical dimension, of particles arising 
from the vacuum, interacting (in a two-dimensional space) 
and finally annihilating one another.
The invariant takes the form of an amplitude for this process 
that is computed through the association of the Yang-Baxter 
solution $R$ as the 
scattering matrix at the crossings and the minima and maxima 
as creation and annihilation operators. Thus we can write the 
amplitude in the form
\[Z_{K} = \langle CUP|M|CAP\rangle \] where $\langle CUP|$ 
denotes the composition of cups, $M$ is the
composition of elementary braiding matrices, and 
$|CAP\rangle $ is the composition of caps. We regard
$\langle CUP|$ as the preparation of this state, and 
$|CAP\rangle $ as the measurement of this state. In order to
view $Z_{K}$ as a quantum computation, $M$ must be a 
unitary operator. This is the case when 
the $R$-matrices (the solutions to the Yang-Baxter
equation used in the model) are unitary. Each $R$-matrix 
is viewed as a a quantum gate (or
possibly a composition of quantum gates), and the 
vacuum-vacuum diagram for the knot is interpreted
as a quantum computer. This quantum computer will 
probabilistically (via quantum amplitudes) compute the values of the states
in the state sum for $Z_{K}$.  

\begin{center}
{\tt    \setlength{\unitlength}{0.92pt}
\begin{picture}(480,243)
\thinlines    \put(34,175){\line(1,0){202}}
              \put(33,97){\line(1,0){202}}
\thicklines   \put(71,97){\circle*{10}}
              \put(190,176){\circle*{10}}
              \put(152,174){\circle*{10}}
              \put(108,175){\circle*{10}}
              \put(150,96){\circle*{10}}
              \put(191,97){\circle*{10}}
              \put(115,98){\circle*{10}}
              \put(70,175){\circle*{10}}
              \put(149,74){\line(1,0){41}}
              \put(190,73){\line(0,1){28}}
              \put(149,98){\line(0,-1){23}}
              \put(115,77){\line(0,1){22}}
              \put(70,76){\line(1,0){45}}
              \put(70,96){\line(0,-1){20}}
              \put(152,192){\line(0,-1){19}}
              \put(110,193){\line(1,0){42}}
              \put(109,175){\line(0,1){18}}
\thinlines    \put(79,1){\framebox(283,43){$Z_{K}=\langle CAP|M|CUP\rangle $}}
              \put(209,120){\framebox(39,42){M}}
              \put(277,101){\framebox(151,37){Unitary Braiding}}
              \put(269,138){\framebox(166,37){Quantum  Computation}}
              \put(254,176){\makebox(200,43){$\langle CAP| (Measurement)$}}
              \put(255,66){\makebox(224,36){$|CUP\rangle  (Preparation)$}}
              \put(1,63){\vector(1,0){261}}
              \put(16,59){\vector(0,1){183}}
\thicklines   \put(190,218){\line(-1,0){40}}
              \put(190,178){\line(0,1){40}}
              \put(69,218){\line(1,0){41}}
              \put(69,177){\line(0,1){41}}
              \put(190,138){\line(0,-1){42}}
              \put(70,136){\line(0,-1){41}}
              \put(110,218){\line(1,0){40}}
              \put(149,98){\line(-1,1){16}}
              \put(110,135){\line(1,-1){16}}
              \put(152,138){\line(-1,-1){38}}
              \put(150,177){\line(1,-1){40}}
              \put(190,178){\line(-1,-1){16}}
              \put(151,138){\line(1,1){16}}
              \put(70,136){\line(1,1){16}}
              \put(110,176){\line(-1,-1){16}}
              \put(70,175){\line(1,-1){40}}
\end{picture}}

{\bf Figure 12  A Knot Quantum Computer} 
\end{center}
\vspace{3mm}

The form of the model proposed for translating 
the Jones polynomial to a quantum 
computation is also the form of models for
anyonic quantum computation (See 
\cite{F, FR98, FLZ, Freedman5, Freedman6}). 
In an anyonic model, the braiding corresponds to the motion of
configurations of particles in a two-dimensional 
space.  These theories are directly related to 
quantum link invariants and to topological quantum
field theories \cite{WIT}. It is hoped that quantum 
computing placed in the anyonic context can be 
made resistant to the effects of decoherence due, in part, to the invariance of
topological structures under perturbation.  
\bigbreak

The formalism of configuring a computation in terms of preparation and measurement in the pattern of Figure 12
can be used in very general quantum computational contexts. For example, let $U$ be a unitary transformation on
$H = V^{\otimes n}$ where $V$ is the complex two-dimesional space for a single qubit. Represent $U$ as a box with
$n$ input lines at the bottom and $n$ output lines at the top, each line corresponding to a single qubit in an
element of the tensor product $H$ with basis $\{|\alpha\rangle | \alpha \,\, \mbox{is a binary string of length} \, n \}.$ 
Let $|\delta\rangle = \Sigma_{\alpha} |\alpha,\alpha\rangle \in H \otimes H$ where $\alpha$ runs over all binary strings of length
$n.$ Note that $\langle \delta|$ is the following covector mapping $H \otimes H$ to the complex numbers $C:$
$$\langle\delta|\alpha, \beta\rangle = 1 \,\, \mbox{if} \,\, \alpha = \beta \,\, \mbox{and} \,\, \langle\delta|\alpha, \beta\rangle = 0
\,\,
\mbox{otherwise.}$$ Now let $W = U \otimes I_{H},$ where $I_{H}$ denotes the identity transformation of $H$ to $H.$
Then $$\langle\delta|W|\delta\rangle = \langle\delta|U\otimes I_{H}|\delta\rangle = $$ 
$$\langle\delta| \Sigma_{\gamma} U^{\gamma}_{\alpha} |\gamma, \alpha\rangle = \Sigma_{\alpha} U^{\alpha}_{\alpha} = tr(U).$$
For example, $\langle\delta|\delta\rangle = 2^{n} = tr(I_{H}).$ See Figure 13 for an illustration of this process.
\bigbreak

{\tt    \setlength{\unitlength}{0.92pt}
\begin{picture}(334,247)
\thinlines    \put(39,105){\makebox(49,39){$U$}}
              \put(232,105){\makebox(101,40){$U \otimes I_{H}$}}
              \put(221,204){\makebox(81,40){$\langle\delta|$}}
              \put(220,1){\makebox(82,43){$|\delta\rangle$}}
              \put(1,69){\dashbox{6}(210,114){}}
              \put(190,4){\line(-1,0){149}}
              \put(190,243){\line(0,-1){239}}
              \put(39,244){\line(1,0){151}}
              \put(158,24){\line(-1,0){97}}
              \put(159,222){\line(0,-1){198}}
              \put(61,223){\line(1,0){98}}
              \put(124,44){\line(-1,0){44}}
              \put(123,204){\line(0,-1){160}}
              \put(82,204){\line(1,0){41}}
              \put(41,84){\line(0,-1){80}}
              \put(61,84){\line(0,-1){59}}
              \put(80,84){\line(0,-1){41}}
              \put(82,165){\line(0,1){39}}
              \put(61,164){\line(0,1){60}}
              \put(39,163){\line(0,1){81}}
              \put(21,83){\framebox(81,81){}}
\end{picture}}

\begin{center}
{\bf Figure 13 - A quantum process to obtain $|tr(U)|.$} 
\end{center}
\vspace{3mm}

Thus we see that we can, for any unitary matrix
$U,$ produce a quantum computational process with preparation $|\delta\rangle$ and measurement $\langle\delta|$ such that {\em the 
amplitude of this process is the trace of the matrix $U$ divided by $(\sqrt{2})^{n}$.} This means that the corresponding quantum computer
computes the probability associated with this amplitude. This probability is the absolute square of the
amplitude and so the quantum computer will have $|tr(U)|^{2}/2^{n}$ as the probability of success  and hence one can find $|tr(U)|$ by
successive trials. We have  proved the 
\bigbreak

\noindent {\bf Lemma.} With the above notation, the absolute value of the trace of a unitary matrix $U$, $|tr(U)|$,
can be  obtained to any desired degree of accuracy from the quantum computer corresponding to $U \otimes I_{H}$ with preparation
$|\delta\rangle$ and measurement $\langle\delta|,$ where $|\delta\rangle = \Sigma_{\alpha} |\alpha,\alpha\rangle \in H \otimes H.$
\bigbreak

\noindent The proof of the Lemma is in the discussion above its statement.

\section{Unitary Representations and Teleportation}
The formalism we used at the end of the last section to describe the (absolute value) of the trace of a unitary matrix contains a hidden
teleportation. It is the purpose of this section to bring forth that hidden connection. 
\bigbreak

First consider the state 
$$|\delta\rangle = \Sigma_{\alpha} |\alpha,\alpha\rangle \in H \otimes H.$$
from the last section, where $H = V^{\otimes n}$  and $V$ is a single-qubit space. 
One can regard $|\delta\rangle$ as a generalization of the $EPR$ state $\langle 00| + \langle 11|.$
\bigbreak

Let $|\psi \rangle \in H$ be an arbitrary pure state in $H.$ Let $\langle {\cal M}|$ be an abitrary element of the dual of $H \otimes H$
and consider the possibility of a successful measurement via $\langle {\cal M}|$ in the first two tensor factors of 
$$|\psi \rangle |\delta\rangle \in H \otimes H \otimes H.$$ The resulting state from this measurement will be
$$\langle {\cal M}|[|\psi \rangle |\delta\rangle].$$
If 
$$\langle {\cal M}| = \Sigma_{\alpha, \beta} M_{\alpha, \beta}\langle \alpha | \langle \beta |, $$ then
$$\langle {\cal M}|[|\psi \rangle |\delta\rangle] = 
\Sigma_{\alpha, \beta} M_{\alpha, \beta}\langle \alpha | \langle \beta | \Sigma_{\gamma, \lambda}\psi_{\gamma} |\gamma\rangle
|\lambda \rangle |\lambda \rangle$$
$$= \Sigma_{\alpha, \beta} M_{\alpha, \beta}  \Sigma_{\gamma, \lambda}\psi_{\gamma} \langle \alpha |\gamma\rangle
\langle \beta |\lambda \rangle |\lambda \rangle$$
$$= \Sigma_{\alpha,\beta} M_{\alpha, \beta} \psi_{\alpha} |\beta \rangle$$
$$= \Sigma_{\beta} [\Sigma_{\alpha} M_{\alpha, \beta} \psi_{\alpha}] |\beta \rangle$$
$$= \Sigma_{\beta} (M^{T}\psi)_{\beta} |\beta \rangle$$
$$= M^{T} |\psi \rangle.$$
Thus we have proved the 
\bigbreak

\noindent {\bf Teleportation Lemma.} Successful measurement via $\langle {\cal M}|$ in the first two tensor factors of 
$$|\psi \rangle |\delta\rangle \in H \otimes H \otimes H$$ results in the state $M^{T} | \psi \rangle$ where the matrix $M$ represents
the  measurment state $\langle {\cal M}|$ in the sense that $$\langle {\cal M}| = \Sigma_{\alpha, \beta} M_{\alpha, \beta}\langle \alpha |
\langle \beta |,$$ and $M^{T}$ denotes the transpose of the matrix $M.$
\bigbreak

This Lemma contains the key to teleportation. Let $| \psi \rangle$ be a state held by Alice, where Alice and Bob share the 
generalized $EPR$ state $| \delta \rangle.$ Alice measures the combined state $|\psi \rangle |\delta\rangle$ and reports to Bob
that she has succeeded in measuring via $\langle {\cal M}|$ (from some list of shared transformations that they have in common) by a
classical transmission of information. By the Lemma, Bob knows that he now has access to the state $M^{T} |\psi \rangle.$ In this 
generalized version of teleportation, we imagine that Alice and Bob have a shared collection of matrices $M$, each coded by 
a bit-string that can be transmitted in a classical channel. By convention, Alice and Bob might take the zero bit-string to denote
lack of success in measuring in one of the desired matrices. Then Alice can send Bob by the classical channel the information of success
in one of the matrices, or failure. For success, Bob knows the identity of the resulting state without measuring it. See Figure 13.1
for a schematic of this process.
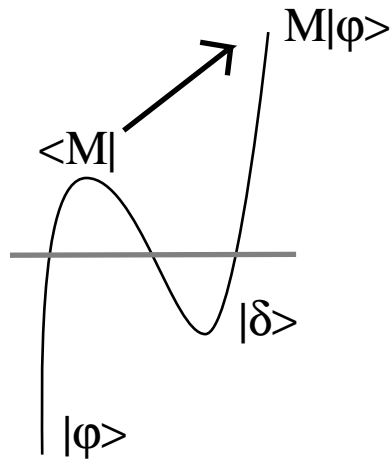
\bigbreak

$$ \picill4inby2.5in(Teleport)  $$

\begin{center}
{\bf Figure 13.1 - Matrix Teleportation}
\end{center}

In the case of success, and if the matrix $M$ is unitary, Bob can apply $(M^{T})^{-1}$ to the transmitted state and know that he now has
the original state $| \psi \rangle$ itself. The usual teleportation scenario, is actually based on a list of unitary transformations
sufficent to form a basis for the measurement states. Lets recall how this comes about.
\bigbreak

First take the case where $M$ is a unitary $2 \times 2$ matrix and let $\sigma_1, \sigma_2, \sigma_3$ be the three Pauli matrices
$$\sigma_{1} =  \left[
\begin{array}{cc}
     1 & 0  \\
     0 & -1
\end{array}
\right] ,\sigma_{2} =  \left[
\begin{array}{cc}
     0 & 1  \\
     1 & 0
\end{array}
\right],\sigma_{2} =  \left[
\begin{array}{cc}
     0 & i  \\
     -i & 0
\end{array}
\right] $$
We replace $\sigma_{3}$ by $-i\sigma_{3}$ (for ease of calculation) and obtain the three matrices $X$, $Y$, $Z:$
$$X =  \left[
\begin{array}{cc}
     1 & 0  \\
     0 & -1
\end{array}
\right], Y=  \left[
\begin{array}{cc}
     0 & 1  \\
     1 & 0
\end{array}
\right],Z =  \left[
\begin{array}{cc}
     0 & 1  \\
    -1 & 0
\end{array}
\right] $$
\bigbreak

\noindent {\bf Basis Lemma.} Let $M$ be a $2 \times 2$ matrix with complex entries. Let the {\em measuring state for $M$} be the
state
$$\langle {\cal M}| = M_{00}|00 \rangle + M_{01}|01 \rangle + M_{10}|10 \rangle + M_{11}|11 \rangle.$$
Let $\langle { \cal XM}|$ denote the measuring state for the matrix $XM$ (similarly for $YM$ and $ZM$).
Then the vectors $$\{ \langle {\cal M}|, \langle { \cal XM}|, \langle {\cal YM}|,\langle { \cal ZM}| \}$$
are orthogonal in the complex vector space $V \otimes V$ if and only if $M$ is a multiple of a unitary matrix $U$ of the form
$$U =  \left[
\begin{array}{cc}
     z & w  \\
     -\bar{w} & \bar{z}
\end{array}
\right]$$ with complex numbers $z$ and $w$ as generating entries.
\bigbreak

\noindent {\bf Proof.} We leave the proof of this Lemma to the reader. It is a straightforward calculation.
\bigbreak

This Lemma contains standard teleportation procedure when one takes $M = I$ to be the identity matrix. Then the four measurement states
$$\{ \langle {\cal I}|, \langle { \cal X}|, \langle {\cal Y}|,\langle { \cal Z}| \}$$ form an orthogonal basis and by the Telportation
Lemma, they successfully transmit $\{ |\psi \rangle, X^{T}|\psi \rangle, Y^{T}|\psi \rangle ,Z^{T}|\psi \rangle \}$ respectively. Bob can
rotate each of  these received states back to $|\psi \rangle$ by a unitary transformation (Remember that states are determined up to
phase.). In this form, the Lemma shows that we can, in fact, teleport any $2 \times 2$ unitary matrix transformation $U.$ We take $M =
U,$ and take the othogonal  basis provided by the Lemma. Then a $2$-qubit classical transmission from Alice to Bob will enable Bob to
identify the  measured state and he can rotate it back to $U|\psi \rangle.$

Note that for $H = V^{\otimes n}$ we can consider the matrices $$T_{\alpha, \beta} = T_{\alpha(1),\beta(1)}\otimes \cdots \otimes
T_{\alpha(n), \beta(n)}$$ where $\alpha = (\alpha(1), \cdots , \alpha(n))$ and $\beta = (\beta(1), \cdots , \beta(n))$ are bit-strings
of length $n$ and 
$T_{0,0} = I, T_{0,1} = X, T_{1,0} = Y, T_{1,1} = Z$ are the modified Pauli matrices discussed above. Then just as in the above Lemma, if
$U$ is a  unitary matrix defined on $H,$ then the set of measurement states $\langle {\cal T_{\alpha,\beta}U}|$ for the matrices
$T_{\alpha,\beta}U$ are an orthogonal basis for $H \otimes H.$ Hence we can teleport the action of the arbitrary unitary matrix $U$ from
Alice to Bob, at the expense of a transmission of $2^{n}$ classical bits of information. This means that, we can construct an
arbitrary unitary transformation (hence an idealized quantum computer) almost entirely by using quantum measurments. This result should
be  compared with the results of \cite{GC}, and \cite{RB}, which we shall do in a paper subsequent to the present work.  If Alice and Bob
conicide as observers, then there is no need to transmit the classical bits. The result of a given measurement is an instruction to
perform one of a preselected collection of unitary transformations on the resulting state. 
\bigbreak

There are a number of lines that we will follow in susequent papers related to the points made in this section. In particular, it is
certainly of interest that one can partially teleport transformations $M$ that are not unitary, at the cost of having only partial
information beforehand of the success of any given measurement. In particular, this means that we could consider computing results 
such as traces or generalized traces of matrices that are not unitary. In this way we could examine computations of knot and link
invariants that are based on  non-unitary solutions to the Yang-Baxter equation. All of this will be the subject of another paper.
In the next section we turn to the subject of quantum computation of link invariants based on unitary solutions to the Yang-Baxter
equation.
\bigbreak

\section{Unitary Representations of the Braid Group and the Corresponding Quantum Computers}

Many questions are raised by the formulation of a quantum computer associated with a given
link diagram configured as preparation, unitary transformation and measurement. Unitary solutions to the
Yang-Baxter equation (or unitary representations of the Artin braid group) that also give link invariants are not so
easy to come by. Here we give a unitary representation that 
computes the Jones polynomial for closures of 3-braids. This  representation provides a test case for the corresponding
quantum computation. We now analyse this representation by making explicit how the bracket polynomial \cite{KA87, KA89, KP, QCJP} is
computed from it.
\vspace{3mm}

The idea behind the construction of this representation depends upon the algebra generated by two single qubit density matrices
(ket-bras).
Let
$|v\rangle$ and 
$|w\rangle$ be two qubits in $V,$ a complex vector space of dimension two over the complex numbers. Let
$P = |v\rangle\langle v|$ and $Q=|w\rangle\langle w|$ be the corresponding ket-bras.  Note that
$$P^2 = |v|^{2}P,$$
$$Q^2 = |w|^{2}Q,$$
$$PQP = |\langle v|w \rangle|^{2}P,$$ 
$$QPQ= |\langle v|w\rangle|^{2}Q.$$
$P$ and $Q$ generate a representation of the Temperley-Lieb algebra \cite{QCJP}. One can adjust parameters to make a representation of
the three-strand braid group in the form
$$s_{1} \longmapsto rP + sI,$$
$$s_{2} \longmapsto tQ + uI,$$
where $I$ is the identity mapping on $V$ and $r,s,t,u$ are suitably chosen scalars. In the following we use this method to adjust
such a representation so that it is unitary. Note that it is possible for the representation to be unitary even though its
mathematical ``parts" $P$ and $Q$ are not unitary. Note also that the resulting representation is made entirely from local unitary 
transformations, so that while there is measurement of topological entanglement, there is no quantum entanglement of any corresponding
quantum states.
\bigbreak

The representation depends on two symmetric but non-unitary matrices $U_{1}$ and $U_{2}$ with

$$U_{1} =  \left[
\begin{array}{cc}
     d & 0  \\
     0 & 0
\end{array}
\right] $$

\noindent and 

$$U_{2} =  \left[
\begin{array}{cc}
     d^{-1} & \sqrt{1-d^{-2}}  \\
         \sqrt{1-d^{-2}}  & d - d^{-1}
\end{array}
\right]. $$
Note that $U_{1}^{2} = dU_{1}$ and $U_{2}^{2} = dU_{1}.$ Moreover, $U_{1}U_{2}U_{1} = U_{1}$ and $U_{2}U_{1}U_{2} = U_{1}.$
 This is an example of a specific
representation of the Temperley-Lieb algebra \cite{KA87, QCJP}.
\noindent The desired representation of the Artin braid group is given on the two braid generators for the three strand braid group by the
equations:

$$\Phi(s_{1})= AI + A^{-1}U_{1},$$
$$\Phi(s_{2})= AI + A^{-1}U_{2}.$$
Here $I$ denotes the $2 \times 2$ identity matrix.

\noindent For any $A$ with $d = -A^{2}-A^{-2}$ these formulas define a representation of the braid group. With 
$A=e^{i\theta}$, we have $d = -2cos(2\theta)$. We find a specific range of
angles $|\theta| \leq \pi/6$ and $|\theta - \pi| \leq \pi/6$ {\it that give unitary representations of
the three-strand braid group.} Thus a specialization of a more general represention of the braid group gives rise to a continuum family
of unitary representations of the braid group.
\vspace{3mm}

Note that $tr(U_{1})=tr(U_{2})= d$ while $tr(U_{1}U_{2}) = tr(U_{2}U_{1}) =1.$
If $b$ is any braid, let $I(b)$ denote the sum of the exponents in the braid word that expresses $b$.
For $b$ a three-strand braid, it follows that 
$$\Phi(b) = A^{I(b)}I + \Pi(b)$$
\noindent where $I$ is the $ 2 \times 2$ identity matrix and $\Pi(b)$ is a sum of products in the Temperley-Lieb algebra 
involving $U_{1}$ and $U_{2}.$ Since the Temperley-Lieb algebra in this dimension is generated by $I$,$U_{1}$, $U_{2}$,
$U_{1}U_{2}$ and $U_{2}U_{1}$, it follows that the value of the bracket polynomial of the closure of the braid $b$, denoted
$<\overline{b}>,$ can be calculated directly from the trace of this representation, except for the part involving the identity matrix. The
result is the equation 
$$<\overline{b}> = A^{I(b)}d^{2} + tr(\Pi(b))$$
\noindent where $\overline{b}$ denotes the standard braid closure of $b$, and the sharp brackets denote the bracket polynomial. 
From this
we see at once that 
$$<\overline{b}> = tr(\Phi(b)) + A^{I(b)}(d^{2} -2).$$

It follows from this calculation that the question of computing the bracket polynomial for the closure of the three-strand
braid $b$ is mathematically equivalent to the problem of computing the trace of the matrix $\Phi(b).$ To what extent can our 
quantum computer determine the trace of this matrix? We have seen just before this subsection that a quantum computation can determine 
the absolute value of the trace by repeated trials. This shows that a major portion of the Jones polynomial for three strand braids can
be done by quantum computation.
\vspace{3mm}

\subsection {The Invariant based on $R$}
A second example is given by the invariant discussed in the previous section.
In that case, we have the formula $$\tau(b) = tr(rep_{n}(b))$$  taken up to multiples of the square root of $2$, and the matrix 
$rep_{n}(b)$ is unitary for any braid $b$ in an $n$-strand braid group for arbitrary positive integer $n.$ This invariant can be
construed as the trace of unitary matrix for a quantum computation. Since, as we have seen, knowledge of the invariant often depends upon
knowing the global sign of the trace of $rep_{n}(b)$, it is not enuough to just compute the absolute value of this trace. Nevertheless
some topological information is available just from the absolute value. 
\bigbreak

\section{ Quantum Entanglement and Topological Entanglement}
The second question about unitary solutions to 
the Yang-Baxter equation is the matter of understanding 
their capabilities in entangling quantum states.
We use the criterion that 
\[
\phi = a|00\rangle  + b|01\rangle  + c|10\rangle  + d|11\rangle 
\]  
is entangled if and only if
$ad - bc \ne 0$.  This criterion is generalized 
to higher dimensional pure states in our papers \cite{TEQE, Spie, L,LO, Lomonaco2}.
\bigbreak
   
\noindent In \cite{QCJP, TEQE, Spie},
we discovered families of unitary solutions to the Yang-Baxter 
equation that detect topological linking if and only 
if the gates corresponding to these solutions can entangle quantum states.
\bigbreak

\noindent Is there a deeper connection between topological entanglement and quantum entanglement?
We believe that more exploration is called for before a definitive answer to this question can be
formulated. We need more bridges between quantum topology and quantum computation.
\bigbreak

\noindent The matrix

\[R = \left( \begin{array}{cccc}
1/\sqrt{2} & 0 & 0  & 1/\sqrt{2}\\
0  & 1/\sqrt{2} & -1/\sqrt{2} & 0\\
0 & 1/\sqrt{2} & 1/\sqrt{2} & 0 \\
-1/\sqrt{2} & 0 & 0 & 1/\sqrt{2}\\
\end{array} \right)\]
is a unitary solution of the Yang-Baxter 
equation; and it is highly entangling for quantum states. It takes
the  standard basis for the tensor product 
of two single qubit spaces to the Bell basis. 
On the topological side, $R$ generates a 
new and non-trivial invariant of knots and links. 
On the quantum side, $R$ is a universal gate at the same level as $CNOT,$ as we showed in 
Theorems 2 and 3.
Thus $R$ is a good example of a transformation that can be examined
fruitfully in both the quantum and the topological contexts.
\bigbreak

\subsection{Linking Numbers and the Matrix $R'.$}

The unitary $R'$ matrix that we have considered in this paper gives rise to a
non-trivial invariant of links. The discussion in this section summarizes our treatement of this 
invariant in \cite{TEQE}. Here we discuss the
invariant associated with the specialization of $R'$ with so that

$$R' = \left[ \begin{array}{cccc} a & 0 & 0 & 0 \\ 0 & 0 & c & 0 \\ 0 & c & 0 & 0
\\ 0 & 0 & 0 & a \end{array} \right].$$

  The invariant is calculated from a state summation associated with the matrix $R'$ and can be shown to have the form 
$$Z_{K} = 2(1 + (c^{2}/a^{2})^{lk(K)})$$  for two-comonent links $K$, where
$lk(K)$ denotes the linking number of the two components of $K.$ We show that {\em for
this specialization of the $R'$ matrix the operator $R'$ entangles quantum states
exactly when it can detect linking numbers in the topological context.}
\bigbreak

Here is a description of the state sum: Label each component of the diagram with either $0$ or $1$.
Take vertex weights of $a$ or $c$ for each local labelling of a positive crossing as
shown in Figure 14. For a negative crossing 
(obtained by interchanging over-crossing and under-crossing segments at a positive crossing)
the corresponding labels are
$1/a$ and $1/c$ (which are the complex  conjugates of $a$ and $c$ repsectively, when $a$ and $c$ are unit complex numbers).  
Let each state (labeling of the diagram by zeroes and ones) contribute the product of its 
vertex weights. Let $\Sigma(K)$ denote the sum over all the states of the products of the vertex
weights. Then one can verify that $Z(K) = a^{-w(K)} \Sigma(K)$ where $w(K)$ is the sum of the crossing
signs of the diagram $K.$

{\tt    \setlength{\unitlength}{0.92pt}
\begin{picture}(202,203)
\thicklines   \put(173,21){\makebox(15,18){0}}
              \put(27,51){\makebox(15,18){0}}
              \put(26,173){\makebox(15,18){0}}
              \put(56,141){\makebox(15,18){0}}
              \put(142,52){\makebox(19,19){1}}
              \put(143,173){\makebox(19,19){1}}
              \put(176,139){\makebox(19,19){1}}
              \put(56,21){\makebox(19,19){1}}
              \put(59,179){\makebox(23,23){a}}
              \put(181,60){\makebox(20,21){c}}
              \put(61,61){\makebox(20,21){c}}
              \put(178,178){\makebox(23,23){a}}
              \put(120,41){\vector(1,0){79}}
              \put(160,1){\vector(0,1){34}}
              \put(160,49){\vector(0,1){32}}
              \put(2,42){\vector(1,0){79}}
              \put(42,2){\vector(0,1){34}}
              \put(41,50){\vector(0,1){32}}
              \put(121,160){\vector(1,0){79}}
              \put(161,120){\vector(0,1){34}}
              \put(160,168){\vector(0,1){32}}
              \put(40,169){\vector(0,1){32}}
              \put(41,121){\vector(0,1){34}}
              \put(1,161){\vector(1,0){79}}
\end{picture}}

{\bf Figure 14 - Positive Crossing Weights}
\bigbreak

{\tt    \setlength{\unitlength}{0.92pt}
\begin{picture}(290,287)
\thicklines   \put(28,51){\makebox(20,22){0}}
              \put(101,86){\makebox(20,22){0}}
              \put(216,182){\makebox(20,22){0}}
              \put(258,63){\makebox(17,19){1}}
              \put(191,52){\makebox(17,19){1}}
              \put(221,23){\makebox(17,19){1}}
              \put(68,92){\makebox(17,19){1}}
              \put(60,24){\makebox(17,19){1}}
              \put(231,92){\makebox(17,19){1}}
              \put(258,245){\makebox(17,19){1}}
              \put(95,220){\makebox(20,22){0}}
              \put(224,249){\makebox(20,22){0}}
              \put(63,251){\makebox(20,22){0}}
              \put(59,181){\makebox(20,22){0}}
              \put(186,212){\makebox(17,19){1}}
              \put(26,212){\makebox(20,22){0}}
              \put(208,84){\vector(1,0){80}}
              \put(287,83){\vector(0,-1){80}}
              \put(288,3){\vector(-1,0){80}}
              \put(247,45){\vector(0,1){32}}
              \put(247,90){\vector(0,1){34}}
              \put(246,123){\vector(-1,0){81}}
              \put(165,124){\vector(0,-1){79}}
              \put(167,45){\vector(1,0){80}}
              \put(208,3){\vector(0,1){36}}
              \put(208,51){\vector(0,1){32}}
              \put(47,85){\vector(1,0){80}}
              \put(126,84){\vector(0,-1){80}}
              \put(127,4){\vector(-1,0){80}}
              \put(86,46){\vector(0,1){32}}
              \put(86,91){\vector(0,1){34}}
              \put(85,124){\vector(-1,0){81}}
              \put(4,125){\vector(0,-1){79}}
              \put(6,46){\vector(1,0){80}}
              \put(47,4){\vector(0,1){36}}
              \put(47,52){\vector(0,1){32}}
              \put(46,244){\vector(1,0){80}}
              \put(125,243){\vector(0,-1){80}}
              \put(126,163){\vector(-1,0){80}}
              \put(85,205){\vector(0,1){32}}
              \put(85,250){\vector(0,1){34}}
              \put(84,283){\vector(-1,0){81}}
              \put(3,284){\vector(0,-1){79}}
              \put(5,205){\vector(1,0){80}}
              \put(46,163){\vector(0,1){36}}
              \put(46,211){\vector(0,1){32}}
              \put(205,212){\vector(0,1){32}}
              \put(205,164){\vector(0,1){36}}
              \put(164,206){\vector(1,0){80}}
              \put(162,285){\vector(0,-1){79}}
              \put(243,284){\vector(-1,0){81}}
              \put(244,251){\vector(0,1){34}}
              \put(244,206){\vector(0,1){32}}
              \put(285,164){\vector(-1,0){80}}
              \put(284,244){\vector(0,-1){80}}
              \put(205,245){\vector(1,0){80}}
\end{picture}}

{\bf Figure 15  - Zero-One States for the Hopf Link} \bigbreak

For example, view Figure 15. Here we show the zero-one states for the Hopf link $H$.
The $00$ and $11$ states each contribute $a^2,$ while the $01$ and $10$ states contribute
$c^2.$ Hence $\Sigma(H) = 2(a^{2} + c^{2})$ and $$a^{-w(H)}\Sigma(H) = 
2(1 + (c^{2}/a^{2})^{1}) = 2(1 + (c^{2}/a^{2})^{lk(H)}),$$ as expected.
\bigbreak

The 
calculation of the invariant in this form is an analysis of quantum networks with cycles in the
underlying graph. In this form of calculation we are concerned with those states of the network
that correspond to labelings by qubits that are compatible with the entire network structure. One considers only
quantum states that are compatible with the interconnectedness of the network as a whole.
\smallbreak

\subsection {The Question About Invariants and Entanglement}
We have seen that there are examples, such as the one given above, where topological entanglement measures, and measures of quantum 
entanglement are related to one another. In that example we found the the solution $R'$ to the Yang-Baxter equation would,
as an operator on states, entangle quantum states exactly when the invariant could measure linking numbers. We have also discussed  the
invariant assoicated with the universal gate $R$ and shown that it detects many topological situations that are quite subtle. For 
example, it can measure the linkedness of the Borromean rings and the linkedness of the Whitehead link, both of which are situations
where the linking numbers are zero. And yet, we have also given an example, in the previous section, of a representation of the braid
group on three strands, $B_3,$ (not constructed from a solution to the Yang-Baxter equation) that produces the Jones polynomial for
closures of three-stranded braids, but is defined on a single qubit. Since this last representation acts only on one qubit, there is no
entanglement associated with it. Therefore it remains, at this  writing, unclear just what is the relationship between the  quantum
entangling properties of braid group representations and their ability to measure topological entanglement. In a sequel to this paper we
will concentrate this analysis just on invariants assoicated with  solutions to the Yang-Baxter equation.

\subsection{The Aravind Hypothesis}
Link diagrams can be used as graphical devices 
and holders of information. In this vein 
Aravind \cite{Ara} proposed that
the entanglement of a link should correspond to 
the entanglement of a state. {\em Observation of a link 
would be  modeled by deleting one
component of the link.} 
A key example is the 
Borromean rings. See Figures 7 and 16.

\begin{center}
{\tt    \setlength{\unitlength}{0.92pt} \begin{picture}(145,141) \thicklines  
\put(61,138){\line(1,0){81}} \put(142,23){\line(0,1){114}}
\put(122,22){\line(1,0){19}} \put(61,22){\line(1,0){50}}
\put(61,75){\line(0,-1){53}} \put(61,138){\line(0,-1){51}}
\put(116,83){\line(-1,0){11}} \put(116,3){\line(0,1){80}}
\put(3,3){\line(1,0){113}} \put(3,82){\line(0,-1){80}}
\put(13,82){\line(-1,0){10}} \put(28,82){\line(1,0){62}}
\put(21,42){\line(0,1){79}} \put(53,42){\line(-1,0){32}}
\put(99,42){\line(-1,0){34}} \put(102,42){\line(0,0){0}}
\put(101,122){\line(0,-1){80}} \put(69,122){\line(1,0){32}}
\put(21,122){\line(1,0){32}} \end{picture}}

{\bf Figure 16 -  Borromean Rings}
\end{center}

\noindent 
Deleting any component of the Boromean rings 
yields a remaining pair of unlinked rings. The Borromean 
rings are entangled, but any two of them are unentangled.
In this sense the Borromean rings are analogous to 
the $GHZ$ state $|GHZ\rangle  = (1/\sqrt{2})(|000\rangle  + |111\rangle )$.
Observation in any factor of the $GHZ$ yields 
an unentangled state. Aravind points out that 
this property is basis dependent. {\em We point out 
that there are states whose entanglement 
after an observation is a matter of probability (via quantum amplitudes).}
Consider for example the state 

{\bf \[
|\psi \rangle=(1/2)(|000\rangle  + |001\rangle  + |101\rangle  + |110\rangle ).
\]} 

\noindent Observation in any coordinate yields an entangled or an
unentangled state with equal probability. For example

{\bf \[
|\psi \rangle=(1/2)(|0\rangle(|00\rangle  + |01\rangle)  + |1\rangle(|01\rangle  + |10\rangle) )
\] }

\noindent so that projecting to $|0\rangle$ in the first coordinate yields an unentangled state, while projecting to $|1\rangle$ yields an
entangled state, each with equal probability.
\bigbreak
 
New ways to use link diagrams must be invented to map the properties 
of such states. {\em We take
seriously the problem of classifying the topological 
entanglement patterns of quantum states.} We are convinced 
that such a classification will be of
practical importance to quantum computing, distributed 
quantum computing and relations with quantum information protocols.
\bigbreak

\section{Braiding and Topological Quantum Field Theory}
The purpose of this section is to discuss in a very general way how braiding is related to topological quantum field theory
and to the enterprise \cite{Freedman5} of using this sort of theory as a model for anyonic quantum computation. The ideas in the subject
of topological quantum field theory are well expressed in the book \cite{AT} by Michael Atiyah and the paper \cite{WIT} by Edward Witten.
The simplest  case of this idea is C. N. Yang's original interpretation of the Yang-Baxter Equation \cite{Yang}.  Yang articulated a
quantum field theory in one dimension of space and one dimension of time in which the $R$-matrix (meaning here any matrix satisfying the
Yang-Baxter equation) was regarded as giving the scattering ampitudes for an interaction of two particles whose (let us say) spins
corresponded  to the matrix indices so that $R^{cd}_{ab}$ is the amplitude for particles of spin $a$ and spin $b$ to interact and produce
particles of spin $c$ and $d.$ Since these interactions are between particles in a line, one takes the convention that the particle with
spin
$a$ is to the left of the particle with spin $b,$ and the particle with spin $c$ is to the left of the particle with spin $d.$
If one follows the braiding diagram for a concatenation of such interactions, then there is an underlying permutation that is obtained
by following the braid strands from the bottom to the top of the diagram (thinking of time as moving up the page). Yang designed the 
Yang-Baxter equation so that {\em the amplitudes for a composite process depend only on the underlying permutation corresponding to the
process and not on the individual sequences of interactions.} The simplest example of this is the diagram for the Yang-Baxter equation 
itself as we have shown it in Figure 1.
\bigbreak

In taking over the Yang-Baxter equation for topological purposes, we can use the same intepretation, but think of the diagrams with 
their under- and over-crossings as modeling events in a spacetime with two dimensions of space and one dimension of time. The extra
spatial dimension is taken in displacing the woven strands perpendicular to the page, and allows us to use both braiding operators $R$ and
$R^{-1}$ as  scattering matrices. Taking this picture to heart, one can add other particle properties to the idealized theory. In
particular one can  add fusion and creation vertices where in fusion two particles interact to become a single particle and in creation
one particle  changes (decays) into two particles. Matrix elements corresponding to trivalent vertices can represent these interactions.
See Figure 17.
\bigbreak

$$ \picill3inby1.5in(CreationFusion)  $$

\begin{center}
{\bf Figure 17 -Creation and Fusion}
\end{center}

Once one introduces trivalent vertices for fusion and creation, there is the question how these interactions will behave in respect to 
the braiding operators. There will be a matrix expression for the compositions of braiding and fusion or creation as indicated in Figure
18. Here we  will restrict ourselves to showing the diagrammatics with the intent of giving the reader a flavor of these
structures. It is natural to assume that braiding intertwines with creation as shown in Figure 19 (similarly with fusion). This
intertwining identity is clearly the sort of thing that a topologist will love, since it indicates that the diagrams can be interpreted
as embeddings of graphs in three-dimensional space. Thus the intertwining identity is an assumption like the Yang-Baxter equation itself,
that simplifies the mathematical structure of the model.
\bigbreak

$$ \picill3inby1.5in(Braiding)  $$

\begin{center}
{\bf Figure 18 - Braiding}
\end{center}

$$ \picill3inby1.5in(Intertwine)  $$

\begin{center}
{\bf Figure 19 - Intertwining}
\end{center}

It is to be expected that there will be an operator that expresses the recoupling of vertex interactions as shown in Figure 20 and labeled
by $Q.$ The actual formalism of such an operator will parallel the mathematics of recoupling for angular momentum. See for example 
\cite{KL}. If one just considers the abstract structure of recoupling then one sees that for trees with four branches (each with a single
root) there is a cycle of length five as shown in Figure 21. One can start with any pattern of three vertex interactions and 
go through a sequence of five recouplings that bring one back to the same tree from which one started. {\em It is a natural simplifying 
axiom to assume that this composition is the identity mapping.} This axiom is called the {\em pentagon identity}. 
\bigbreak

$$ \picill3inby1.5in(Recouple)  $$

\begin{center}
{\bf Figure 20 - Recoupling}
\end{center}

$$ \picill3inby2.7in(Pentagon)  $$

\begin{center}
{\bf Figure 21 - Pentagon Identity}
\end{center}

Finally there is a hexagonal cycle of interactions between braiding, recoupling and the intertwining identity as shown in Figure 22.
One says that the interactions satisfy the {\em hexagon identity} if this composition is the identity.
\bigbreak

$$ \picill3inby4.5in(Hexagon)  $$

\begin{center}
{\bf Figure 22 - Hexagon Identity}
\end{center}

A {\em three-dimensional topological quantum field theory} is an algebra of interactions that satisfies the Yang-Baxter equation, the
intertwining identity, the pentagon identity  and the hexagon identity. There is not room in this summary to detail the remarkable way
that these properties fit into the topology of knots and three-dimensional manifolds.
As the reader can see, a three dimensional $TQFT$ is a highly simplified theory of point particle interactions in $2+1$
dimensional spacetime. It can be used to articulate invariants of knots and links and invariants of three manifolds. The reader
interested in the
$SU(2)$ case of this structure and its implications for invariants of knots and three manifolds can consult \cite{KL, KP, Kohno,
Crane, MS}. One expects that physical situations involving
$2+1$ spacetime will be approximated by such an idealized theory. It is thought for example, that aspects of the quantum Hall effect
will be related to topological quantum field theory
\cite{Wilczek}. One can imagine a physics where the space is two dimensional and the braiding of  particles corresponds to their exchanges
as though circulating around one another in the plane. Such particles that, unlike fermions, do not just change the amplitude by a sign
under interchange, but rather by a complex phase or even a linear combination of states are called {\em anyons}.  It is hoped that $TQFT$
models will describe applicable physics. One can think about the  possible applications of anyons to quantum
computing. The $TQFT's$ then provide a class of anyonic models where the braiding is essential to the physics and to the
quantum computation. We have given a sketch of this approach here to give the reader a picture of one of the possibilities
of using braiding in quantum computing.  
\bigbreak

$$ \picill3inby3.5in(BraidingOperator)  $$

\begin{center}
{\bf Figure 23 - A More Complex Braiding Operator}
\end{center}
\bigbreak

The key point in the application and relationship of $TQFT$ and quantum information theory is, in our opinion, contained in the 
structure illustrated in Figure 23. There we show a more complex braiding operator, based on the composition of recoupling with the
elementary braiding at a vertex. (This structure is implicit in the Hexagon identity of Figure 22.) The new braiding operator is a 
source of unitary representations of braid group in situations (which exist) where the recoupling transformations are themselves 
unitary. This kind of pattern is implicitly utilized in the work of Freedman and collaborators \cite{F, FR98, FLZ, Freedman5, Freedman6}
and in the case of classical angular momentum formalism has been dubbed a "spin-network quantum simlator" by Rasetti and collaborators
\cite{MR}.
\bigbreak

\section{\bf Discussion}
It is natural to expect relationships between topology and quantum mechanics. For example, Dirac \cite{D} described the
relationship between an observer and a fermion by using the properties of twisted belts  embedded in three dimensional space. These
properties vividly portray the consequences of the fact that $SU(2)$  double covers $SO(3).$ The rotation group $SO(3)$ and the unitary
group $SU(2)$ are involved since a rotation  of the observer is mapped to a unitary transformation of the wave function. The topology of
the belt gives a direct way to  image the properties of this connection, with one full rotation changing the sign of the wave function,
while two full rotations  do not change that sign. In the topological picuture, that relationship between one object and another object
rotated relative to the first object is depicted by a belt connecting them. Topological properties of the belt mimic the
orientation - entanglement relation.
\bigbreak

How might such relationships between topology and quantum mechanics impinge upon quantum computing? The Dirac string trick
suggests that topology may enter in the structure of non-locality and entanglement. On the quantum computing side,
we know many uses for entangled states (e.g. teleportation protocols); and one wants to understand the role of entanglement in the
efficiency of computing procedures. Entanglement in quantum mechanics and entanglement (linking and knotting) in topology can be
related in a number of ways that give rise to a host of research questions. 
\bigbreak

We would like to state some general properties of this quest for relationship between
topology and quantum mechanics: It is normally assumed that one is given the background space over which quantum mechanics appears.
In fact, it is the already given nature of this space that can make non-locality appear mysterious. In writing $|\phi\rangle =
(|01\rangle + |10\rangle)/\sqrt{2},$ we indicate the  entangled nature of this quantum state without giving any hint about the spatial
separation of the qubits that generate the first and second factors of the tensor product for the state. This split between the
properties of the background space and properties of the quantum states is an artifact of the rarefied form given to the algebraic
description of states, but it also points out that it is exactly the separation properties of the topology on the background
space that are implicated in a discussion of non-locality.
\bigbreak

Einstein, Podolsky and Rosen might have argued that if two points in
space are separated by disjoint open sets containing them, then they should behave as though physically independent. Such a postulate
of locality is really a postulate about the relationship of quantum mechanics to the topology of the background space. The Dirac
string trick can be understood in a similar manner. In this way, we see that discussions of non-locality in quantum mechanics are in
fact  discussions of the relationship between properties of the quantum states and properties of the topology of the background
space. Subtle questions related to metric and change of metric give rise to the well-known problems of quantum gravity (since general
relativity must take into account the subtleties of the spacetime metric and the topology of spacetime). 
\bigbreak
 
Approaches such as Roger Penrose's spin networks and the more recent work of John Baez, John Barrett, Louis Crane
Lee Smolin, and others
suggest that spacetime structure should emerge from networks of quantum interactions occurring in a pregeometric or process phase of 
physicality. In such a spin network model, there would be no separation between topological properties and quantum properties.
We intend to carry this discussion to the spin network or to the spin foam level. It is our aim to
deepen the discussion of topology and quantum computing to a level where this can be done in a uniform manner.
\bigbreak

The spin network level is already active in topological models
such as the Jones polynomial, the so called quantum invariants of knots, links and three-manifolds, topological quantum 
field theories \cite{AT, WIT}, and related anyonic models for quantum computing \cite{F, FR98, FLZ}. For example, the bracket model
\cite{KA87,KA89, KP, QCJP}  for the Jones polynomial can be realized by generalization of the Penrose $SU(2)$ spin nets to the
quantum group $SU(2)_q.$ 
\bigbreak

Since  the advent of knot invariants such as the Jones polynomial, spin network studies have involved $q$-deformations of
classical spin networks and the corresponding topological properties. These $q$-deformations are, in turn, directly related to
properties of $q$-deformed Lie algebras (quantum groups, Hopf algebras) containing solutions to the Yang-Baxter equation. Solutions
to the Yang-Baxter equation are maps $R:V \otimes V \longrightarrow V \otimes V$ on the tensor product of two vector spaces  that
represent topological braiding. 
\bigbreak

A direct question important for us is the determination of unitary solutions to
the Yang-Baxter equation, and the investigation of both their topological properties and their quantum information properties.
For the latter we want to know what role such solutions (matrices) can play in quantum computing. Specific questions are how such a
matrix can be used in a quantum computational model for a link invariant, and can the matrix in question map unentangled states to
entangled states. Some of these specific phenomena have been discussed in this paper.
\bigbreak

\end{document}